\documentclass[%
 reprint,
 amsmath,amssymb,
 aps,
]{revtex4-2}

\usepackage{graphicx}
\usepackage{dcolumn}
\usepackage{bm}
\usepackage{braket}
\usepackage{xcolor}
\newcommand\ehat{\mathbf{\hat{e}}}
\newcommand\eqnref[1]{\text{Eq.~}\eqref{#1}}
\begin{document}

\title{Crystalline topological defects within response theory}
\author{Sami Hakani}
    \email{shakani3@gatech.edu}
\author{Itamar Kimchi}%
\email{ikimchi3@gatech.edu}
\affiliation{School of Physics, Georgia Institute of Technology, Atlanta, GA 30332}

\date{November 16, 2023}%

\begin{abstract}
Crystal defects can highlight interesting quantum features by coupling to the low-energy Hamiltonian $H$. Here we show that independently of this $H$ coupling, topological crystalline defects can generate new features by directly modifying the response theory of electric field probes such as Raman scattering. To show this we consider an antiferromagnetic spin-1/2 model $H_{spin}$ on a zigzag chain. Crystalline domain walls between two zigzag domains appear as at most local defects in $H_{spin}$, but as topological (not locally creatable) defects in the Raman operator $R$ of inelastic photon scattering. Using time evolving block decimation (TEBD) numerics, mean field, and bosonization, we show that a finite density of crystalline domain walls shifts the entire Raman signal to produce an effective gap. This lattice-defect-induced Raman gap closes and reopens in applied magnetic fields. We discuss the effect in terms of photons sensing the lattice defects within $R$ as spin-dimerization domain walls, with $Z_2$ character, and a resulting shift of the probed wavevector from $q=0$ to $\pi+\delta q$, giving an $\textit{O}(1)$ change in contrast to local defects. The magneto-Raman singularity from topological lattice defects here relies on the $H_{spin}$ spinon liquid state, suggesting future applications using lattice topological defects to modify response-theory operators independently of $H$ and thereby generate new probes of quantum phases. 
\end{abstract}

\maketitle


\section{Introduction}

Crystalline defects have seen increasing attention for their potential to serve as probes of quantum effects in quantum phases of matter.
Examples of such effects are now known across various types of quantum phases. 
For example, weak 3D topological insulators host a 1D topological modes in dislocation lines \cite{Ran2009}; 
spin-1/2 two leg ladders show singlet breaking with few percent impurities \cite{Azuma1997}; 
defects in quantum spin liquids host low energy modes \cite{Willans2010, Kao2021}. In these and many other examples, crystal or lattice defects probe a quantum phase by modifying the wave function of low energy degrees of freedom or coupling to low energy excitations. In this work we shift the focus away from the question of how defects change any particular low energy quantum state, and instead introduce a different question: Can some types of crystal defects modify the results of certain experimental measurements at low energies, even when they do not modify the low energy state itself?

We answer the question in the positive by defining and analyzing a simple model that shows exactly such an effect. The general principle associated with the effect is that topological defects in the crystal (lattice) order can couple directly to electric fields in experiments with photons or other sources of electric fields. The defects thereby directly modify the probe theory operator, independently of whether or not the defects modify the original system's low energy Hamiltonian. 
We find that these modifications of the response theory operator can produce features in the measurement results that are different in significant
ways from the defect-free case.

In particular we require \textit{topological} crystalline defects. For concreteness we focus on crystalline domain walls of the two zigzag/zag-zig domains of a zigzag chain (a.k.a.\ sawtooth chain with mirror symmetries) of a spin-1/2 antiferromagnet. Such defects may appear in the spin Hamiltonian, but if so, only as local defects (i.e.\ locally creatable, in contrast to topological defects which are not locally creatable). In contrast they can appear in electric field probes, such as the Raman operator of inelastic photon scattering, as true topological defects. As we discuss below, this allows the topological crystalline defects to shift the effective wavevector probed by the photons in a manner related to the defect concentration, resulting in an additive contribution in the probe operator that is $\textit{O}(1)$ for any defect concentration. 
This $\textit{O}(1)$  effect should be contrasted with the additive effect of local defects which for defect concentration $\epsilon$ is only $\textit{O}(\epsilon)$. This is the case even when local defects give qualitatively new features, as when locally reduced symmetry near a defect allows transitions from previously-forbidden matrix elements \cite{Eckmann2012}; these transitions appear with amplitude $\epsilon$.
The $\textit{O}(1)$ additive effects seen here, occurring in a bulk probe even at small defect concentrations, require each defect to modify the probe operator on infinitely many sites. This implies defects that are not locally creatable, i.e.\ topological crystalline defects.

Below we show that the Raman operator for a zigzag domain has explicit spin dimerization; and moreover, zigzag domain walls manifest as dimerization domain walls. It is interesting to consider what such dimerization domain walls would do if they appeared in a Hamiltonian. Previous studies have shown that dimerization domain walls carry a spin-1/2 degree of freedom for protected ``topological'' reasons associated with the quantum anomaly of a single spin-half, with similar effects occurring in 2D frustrated magnets \cite{kimchi2018}. Given that dimerization domain walls in a Hamiltonian carry such topologically-protected spin-1/2 modes that respond to magnetic fields, it is interesting to study the dimerization domain walls in the Raman operator, induced by crystal defects, and ask whether they too have a magnetic response. Surprisingly, below we show that non-magnetic crystal defects indeed  do give rise to a magnetic response within Raman scattering.

If they are found in experiments on a new unstudied material, the surprising effects we predict -- such as a Raman gap closing and reopening with applied magnetic fields -- could easily be misinterpreted. They could naively be thought to arise as an intrinsic magnetic signature of some phase, or as effects from magnetic impurities. The framework we present is necessary for a correct interpretation; in this case, as effects from non-magnetic crystal defects combined with 1D-fractionalized spinon excitations. 

Our main purpose in this work is to argue for the general effect of crystalline topological defects modifying response theory operators, and to present a proof-of-principle toy model. This toy model however may still adequately describe some magnetic insulator compounds. Let us suggest some materials. 
Rb$_2$Cu$_2$Mo$_3$O$_{12}$ is a zigzag spin-half chain compound with a singlet ground state and large couplings (ferromagnetic $J_1\approx -138$ K and antiferromagnetic $J_2 \approx 51$ K); a Luttinger Liquid phase may arise in a small applied field \cite{Hase2004, Hayashida2019}.
AgCrP$_2$S$_6$ is a zigzag chain compound where Kramer's doublet spin-$3/2$ magnetic sites interact with a nearest neighbor antiferromagnetic exchange of order 100 K \cite{Selter2021, Mutka1993}.
Copper benzoate \cite{Dender} is a spin-$1/2$ antiferromagnetic chain compound; the Cu-Cu bonds do not zigzag, but the oxygen octahedra coordinating Cu have alternating orientations on successive Cu sites, so that the symmetry is equivalent to a zigzag chain. Coupling of the lattice to electric fields will generate topological defects in the Raman operator in exchange paths above lowest order, i.e.\ beyond the Loudon-Fleury approximation \cite{yang2021}, for example through Cu-O superexchange, which gives the Cu-O vector dotted with photon polarization.
(The magnetic field should be applied in a direction that avoids generating an effective staggered field \cite{Affleck1999}.)
In all these materials, one limiting factor may be the Raman resolution at low frequencies, which must be high enough to resolve effects at frequencies set by typical inverse domain sizes. Nevertheless here resolving low frequency features is helped by the strong inelastic signal at $\omega\rightarrow 0$, which is not present in typical $q=0$ Raman experiments. 

\begin{figure}
    \centering
    \includegraphics[width=0.52\textwidth]{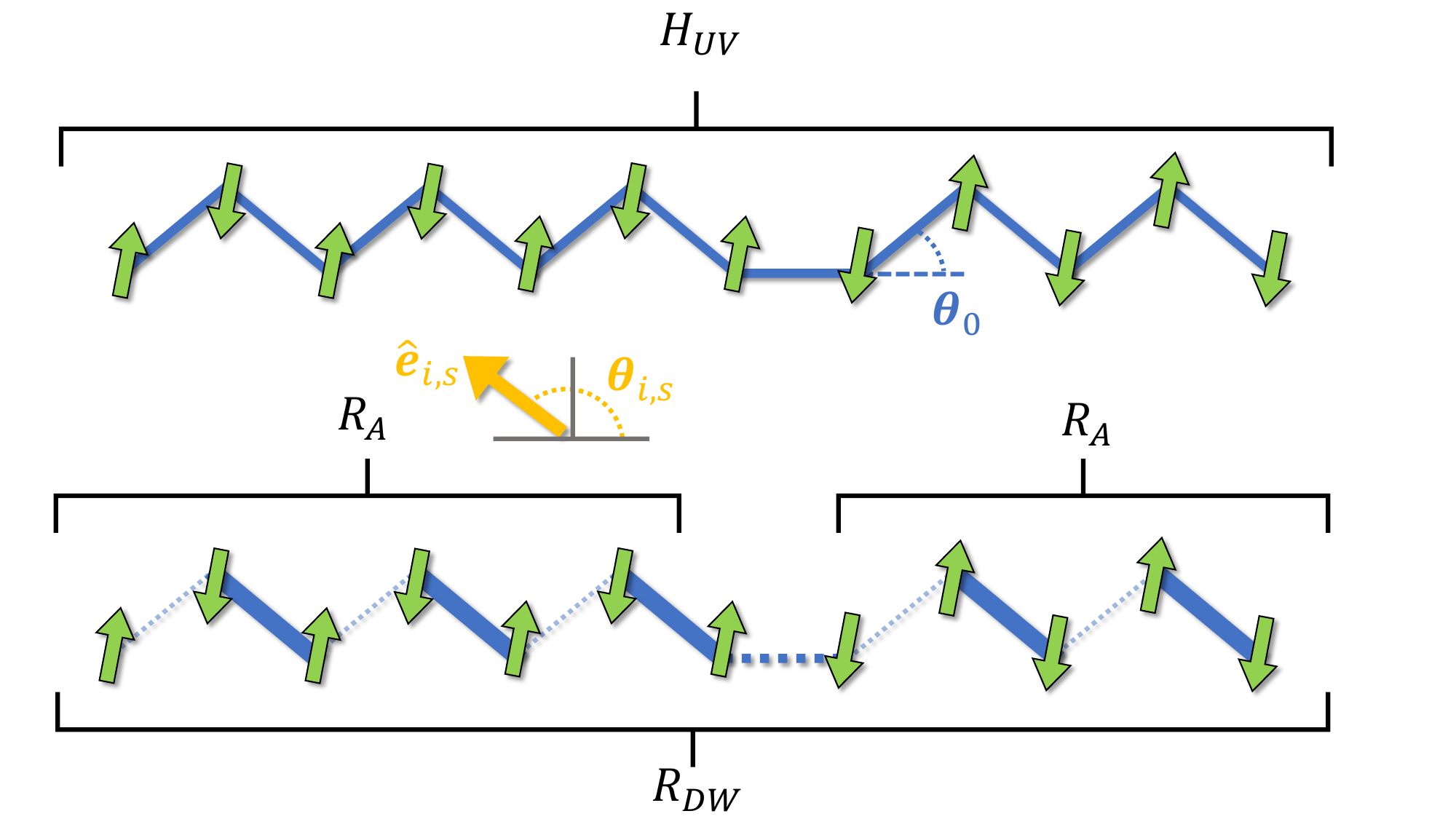}
    \caption{(Top) The lattice-level Hamiltonian $H_{UV}$ for a spin-1/2 zigzag chain contains information on spatial positions. This information may be invisible to the low energy  spin Hamiltonian $H_0$ (Eqn.~\ref{eq:H}) since $H_0$ has identical antiferromagnetic couplings on ``zig'' and ``zag'' bonds. 
    (Bottom) Inelastic Raman scattering  probes the the Raman operator $R$ dynamical correlations within $H_{0}$. However $R$ knows about $H_{UV}$ spatial positions through the photon electric fields. For some photon polarizations $\mathbf{\hat{e}}_{i},\mathbf{\hat{e}}_{s}$ (especially diagonal or cross polarizations so $\theta_i+\theta_s=\pi/2$), the Raman operator is an alternating dimerization operator $R_A$  (Eqn.~\ref{eqn:ra}). 
    Zigzag domain walls in $H_{UV}$, separating the two zigzag domains, do not appear as topological defects within $H_{0}$. However within $R$ they manifest and become dimerization domain walls, separating two distinct $R_A$ dimerization domains. Thus $R_{DW}$ contains a topological defect.}
    \label{fig:zigzagdefects}
\end{figure}

The remainder of this paper is structured as follows. 
In Section \ref{sec:surprises} we present the model (Fig.\ \ref{fig:zigzagdefects}) and discuss its Raman response including in the presence of crystalline domain walls, through the coupling between crystalline domains and photon polarizations (electric fields). For each domain the Raman operator $R$ is a \textit{dimerization} operator which breaks the translation symmetry of the spin Hamiltonian $H$. This is in contrast with spin-Peierls (dimerized or valence bond solid) materials such as CuGeO$_3$, where $R$ and $H$ are both explicitly dimerized \cite{brenig1997, Augier1999}. When Hamiltonians or wavefunctions show explicit dimerization, it is well known that dimerization domain walls carry protected spin-1/2 modes with a strong magnetic signal \cite{kimchi2018}. Surprisingly, here dimerization domain walls in the Raman operator, created by crystalline domain walls, also produce a magnetic signal. Using TEBD (time evolving block decimation) numerics, bosonization, and mean field, we find that a density of domain walls creates a gap in the Raman spectrum. This gap closes and then reopens above a critical magnetic field set by the typical domain size. We explain the effects in terms of a shift of the wavevector probed by the photon scattering, from $q=0$ typical in solid state Raman to $q=\pi$ for a single domain, to $q=\pi+\delta q$ with multiple domains of typical size $\pi/\delta q$, combined with the magnetic field response of the fermionic excitations in the gapless Luttinger liquid phase of a spin-1/2 antiferromagnetic chain. 
  
In Section \ref{sec:discussion} we discuss these results, their restrictions and generalizations.  That these features arise from domain walls rather than local defects is also seen in a $Z_2$ characteristic that we discuss, wherein two defects brought close together have no anomalous response. We discuss the effects of local defects on the 1D spin Hamiltonian, which are known to be RG-relevant in 1D, resulting in a distribution of finite size chain fragments, whose finite size gaps closely mirror the results in the limit discussed above. For both of these limits (unperturbed and fixed-point-fragmented Hamiltonian) we discuss the distribution of domain sizes and its effects on the expected features. Finally we show that the effects here arise from domain walls probing the fermionic spinon excitations, by computing the domain wall Raman response for other candidate phases -- ferromagnets, antiferromagnets, and gapped phases of integer spin chains such as the AKLT phase -- and show that the features described above rely on the gapless spinon phase. Thus the features can also be interpreted as a modification of Raman scattering, via crystalline domain walls, in order to probe the 1D-fractionalized spinon excitations.

\section{Results}\label{sec:surprises}

\subsection{Introduction to Raman Scattering} 
The inelastic Raman scattering spectrum $I(\omega)$ of a system  is specified by its Hamiltonian $H$, the ground state, and a third ingredient, the Raman operator $R$. This third ingredient is needed to describe the photon-in-photon-out scattering process; the wavevector of the photons is sufficiently small that it can be ignored, but photon frequency and polarizations remain important. In terms of $R$,  the spectrum $I(\omega)$ is computed as the dynamical correlation function,
\begin{equation}\label{eqn:spectrum}
    I(\omega) = \frac{1}{2\pi}\int_{-\infty}^\infty dt \ e^{i\omega t} \braket{R(t)R(0)}_0
\end{equation}
Here we work at zero temperature so $\braket{\cdots}_0$ is the ground state expectation value, easily modified for finite temperature.
The key player here is the Raman operator $R$, which includes information on the photon polarizations and associated electric fields. In the simplest limit applicable to a spin model, the so-called Loudon-Fleury (photon-induced superexchange) form of the operator \cite{fleury1968} is
\begin{equation}\label{eqn:fleuryoperator}
    R = \sum_{\textbf{r}_1,\textbf{r}_2} (\mathbf{\hat{e}}_i \cdot \textbf{r}_{12})(\mathbf{\hat{e}}_s \cdot \textbf{r}_{12})A(\textbf{r}_{12}) \textbf{S}_{r_1} \cdot \textbf{S}_{r_2}
\end{equation}
In \eqnref{eqn:fleuryoperator}, $\mathbf{\hat{e}}_{i(s)}$ is the incident (scattered) photon polarization and $\textbf{r}_{12} = \textbf{r}_1 - \textbf{r}_2$. Ratios of $A(\textbf{r}_{12})$ on different bonds are of the order of the ratio of the respective exchange couplings, since they are similarly generated by superexchange. Moreover, from the definitions above it is evident that two Raman operators $R$ and $R'$ yield the same inelastic scattering spectrum under a Hamiltonian $H$ if there exists a real constant $c$ such that $R' = R - cH$, since $cH$ does not time evolve. A pair of $R$ and $R'$ differing by a multiple of $H$ are spectrally equivalent. 

Since photon polarizations ($\mathbf{\hat{e}}_{i,s}$) necessarily couple to spatial degrees of freedom ($\mathbf{r}_{12}$), the Raman operator inherits a rich structure from the spatial geometry of the system, endowing $R$ with symmetries that are different from the symmetries of the Hamiltonian $H$ (Fig.\ \ref{fig:zigzagdefects}). 
    
\subsection{Raman Response of the Clean Zigzag Chain} 
We begin by computing the Raman operator and Raman response for a 1D spin-1/2 zigzag chain in the clean limit, with a single domain. 
This setup was previously considered by Ref.~\cite{sato2012} for a spin system within bosonization at zero magnetic field. 
Here we will consider an effective 1D antiferromagnetic spin-1/2 zigzag chain in the presence of an applied magnetic field in order to highlight the interplay between spatial geometry, Raman operator structure, and the associated Raman response. Importantly, within a single zigzag domain, the spin Hamiltonian $H_{spin}$ for this system is blind to the zigzagging. For simplicity we focus on nearest neighbor couplings and take $H_{spin}$ to be the conventional 1D antiferromagnetic XXZ spin chain given by
\begin{equation}
    H_0 = J\sum_j (\textbf{S}_j \cdot \textbf{S}_{j+1} + (\Delta{-}1) S_j^z S_{j+1}^z) - h^z \sum_j S_j^z \label{eq:H}
\end{equation}
with $J > 0$, and where the $H_0$ subscript refers to the clean limit. Although the ultraviolet Hamiltonian for a spin-1/2 zigzag chain contains information about both spin and lattice degrees of freedom, at low energies the infrared Hamiltonian is an effective 1D spin-1/2 XXZ chain as depicted in Fig.\ \ref{fig:zigzagdefects}. For the remainder of this section, we take the magnetic field $h^z = 0$; magnetic field effects will be considered further below.

The zigzag spin chain geometry and \eqnref{eqn:fleuryoperator} give the following form for the Raman operator when $\Delta = 1$:
\begin{equation}
    R \propto \sum_j \left[ f_{\theta_i,\theta_s,\theta_0} + h_{\theta_i,\theta_s,\theta_0} (-1)^j \right] (\textbf{S}_j \cdot \textbf{S}_{j+1})
\end{equation}
where $\theta_{i,s}$ is the angle of the incident/scattered photon polarization (measured from the zigzag chain axis), $\theta_0$ is the zigzag angle (Fig.\ \ref{fig:zigzagdefects}), $f_{\theta_i,\theta_s,\theta_0} = \cos\theta_i\cos\theta_s \cos^2\theta_0 + \sin\theta_i\sin\theta_s\sin^2\theta_0$, and 
\begin{equation}
h_{\theta_i,\theta_s,\theta_0} = \frac12 \sin(2\theta_0)\sin(\theta_i + \theta_s). 
\end{equation}
When $\theta_i + \theta_s = \frac{\pi}{2}$, the alternating part of $R$ is maximized. Physically, this alternating part is maximized, relative to any other terms which can arise for generic Hamiltonians, for the physically relevant  cases of: cross polarizations ($\theta_i = 0$, $\theta_s = \frac{\pi}{2}$; or vice versa);  and parallel polarizations along a diagonal ($\theta_i = \theta_s = \frac{\pi}{4}$).

When the alternating part of $R$ is considered, the Raman operator for this system has a dimerization structure and is given by
\begin{equation}\label{eqn:ra}
    R_A \propto \sum_j (1+(-1)^j) (\textbf{S}_j \cdot \textbf{S}_{j+1})
\end{equation}
where we recall that Raman operators $R$ and $R'$ are spectrally equivalent if they differ by a multiple of $H$. 
Unlike the spin Hamiltonian which has discrete translational invariance by one lattice site, $R_A$ explicitly breaks this translational invariance. In particular $R_A$ generally also has lower symmetry than the full Hamiltonian of the material: for example, the zigzag chain of Fig.\ \ref{fig:zigzagdefects} has mirror symmetry across each site, while $R_A$ does not. Its symmetry is reduced through the coupling of the crystal lattice with photon polarizations or  electric fields.

While typically a Raman response in solids does not give a gapless inelastic ($\omega \to 0$) response, here at $T = 0$ we capture a gapless inelastic Raman response due to the alternating structure of $R_A$ induced by the zigzag geometry. The lowest order nonvanishing contribution to the Raman response at the mean field level (see below and Appendix \ref{sec:spinonmft}) is given by
\begin{equation}\label{eqn:cleanmf}
    I(\omega) \propto \sqrt{1 - \left( \frac{\omega}{2v_s} \right)^2}
\end{equation}
This continuum is computed at mean field as a 2-particle response. The low energy effective field theory also predicts a gapless inelastic response for $0 \leq \Delta < 1$ \cite{sato2012}. We note, however, at the SU(2) symmetric point ($\Delta = 1$)  bosonization (see below) gives a purely elastic response: $I(\omega) \to \delta(\omega)$. Nonetheless, away from that point, both mean field and  bosonization capture a gapless inelastic Raman response for $R_A$ due to its alternating structure.

\subsection{Raman Response of Zigzag Chains with Zigzag Domain Walls} 

We now turn to Raman scattering in the presence of crystalline topological defects, here zigzag domain walls i.e.\ multiple zigzag domains. Although zigzag domain walls modify the Hamiltonian of the system by acting as a local (not topological) defect, and similarly the local defect part of the domain wall (e.g.\ the horizontal bond) can appear in $R_{DW}$ in a non-universal manner, we address these modifications later and find they do not qualitatively change our conclusions. For now, let us suppose zigzag domain walls only alter the Raman operator, and only by altering each domain.

\subsubsection{Numerics (DMRG and TEBD)}
\begin{figure*} 
    \centering
    \includegraphics[width=0.49\textwidth]{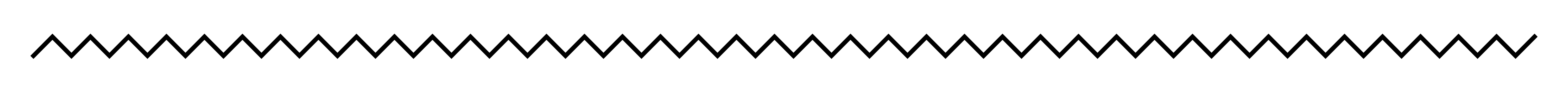}
    \includegraphics[width=0.49\textwidth]{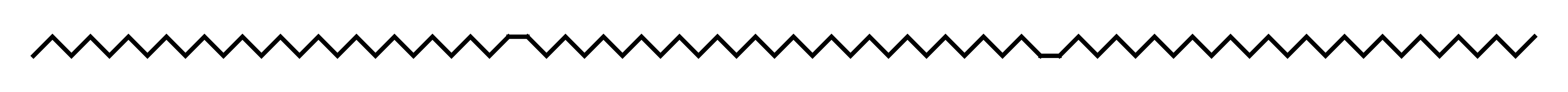} 
    \includegraphics[width=\textwidth]{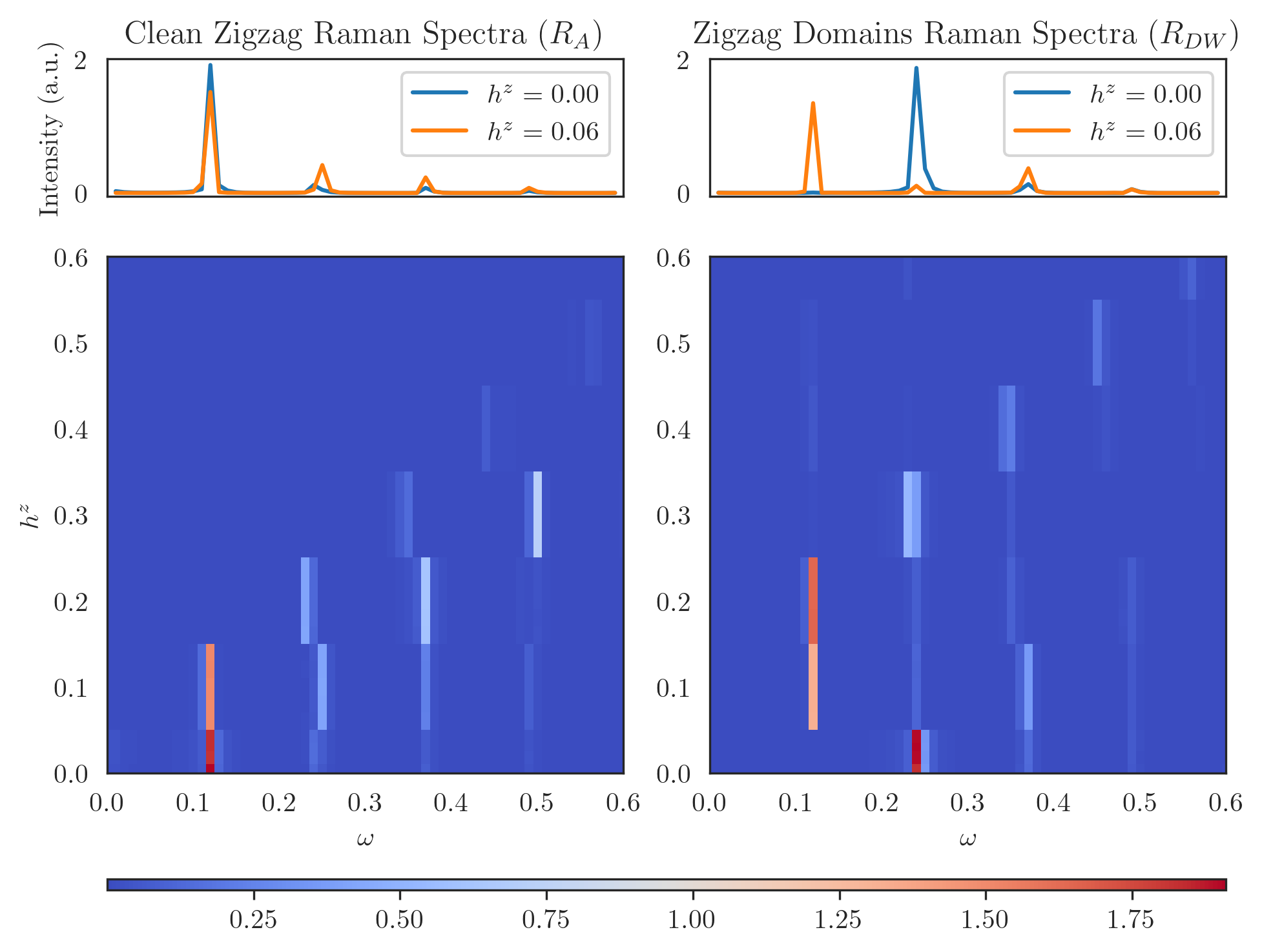}
    \caption{(Top) 
    Inelastic magneto-Raman spectra as a function of frequency $\omega$ and applied magnetic field $h^z$, computed numerically using DMRG and TEBD for a finite size $L{=}80$ Heisenberg spin-1/2 chain ($J{=}1, \Delta{=}1$, open boundaries). Left column: clean system with single $R_A$ domain; right column: three $R_A$ domains in the $R_{DW}$ Raman operator, separated by two domain walls (at $j_1{=}26$ and $j_2{=}54$, giving domain size $L_d{=}27$). In both cases the Hamiltonian $H_0$ is identical; the only difference is in the Raman operator. The line drawing of the corresponding $H_{UV}$ zigzag chain is shown at top. Upper panel: line cuts of spectra at zero and nonzero fields show that while the clean $R_A$ is independent of small fields, the domain walls of $R_{DW}$ show a singular response to magnetic fields. Lower panel: color plots of spectra as function of magnetic field show the full behavior. 
    At zero field, the presence of domain walls shifts the finite size gap to higher energy. This shift persists up to a small critical field; the critical field value is consistent with the theoretical expectation $h_c = \frac12 v_s (\pi/L_d)$ with $v_s = \pi/2$. Spectral weight shifts upward with field above $h^z = 0$ for zero defects and above $h^z = h_c$ for two defects.}
    \label{fig:tebd}
\end{figure*}

Using a combination of denstiy matrix renormalization group (DMRG) and time evolving block decimation (TEBD), we numerically compute the $T = 0$ inelastic Raman scattering spectra as a function of frequency and applied magnetic field, for both zero domain walls and two domain walls (Fig.\ \ref{fig:tebd}). We use spin chains of length $L{=}80$ with open boundary conditions.

Our numerics for the zero domain wall case show a prominent low energy excitation at a  frequency $\omega^*$. In applied field $h^z$, we find $\omega^*$ shifts to higher frequencies linearly with $h^z$. The finite gap at low frequencies can be attributed to the finite size of the system. Indeed we find $(\omega^* / J) \sim L^{-1}$ as the system size is increased.

The two domain wall case shows a zero field gap that is larger than the finite size gap of the zero domain wall case. Moreover, for small applied fields the gap shrinks to the finite size gap of $\omega^*$. At larger fields, the gap increases. 

\subsubsection{Mean field theory} 
To understand these numerical results, let us discuss the results from a mean field computation of the effects of domain walls on $R$. For $\Delta = 0$, the Hamiltonian is an  $XY$ model and the mean field mapping is exact with Jordan-Wigner fermions. Within mean field, the Hamiltonian is a spinon Fermi sea $H_0^{MF} = \sum_k \epsilon_k c_k^\dagger c_k$ with spinon field operators defined by $\{c_k, c_{k'}^\dagger\} = \delta_{kk'}$, dispersion $\epsilon_k = -v_s \cos(k)$, and spinon velocity $v_s$, with $v_s=J$ at $\Delta = 0$ (we take unit lattice spacing throughout). In terms of fermions, the Raman operator $R_A$ is a $q = \pi$ spinon density excitation
\begin{equation}
    R_A = \sum_k V_{k,q=\pi} c_k^\dagger c_{k-\pi}
\end{equation}
where the vertex is $V_{kq} = \frac12 (e^{i(k-q)} + e^{-ik} - 1 - e^{-iq})$.  (Details are below and, including XXZ anisotropy for $R$, in Appendix \ref{sec:spinonmft} and  \ref{sec:linearmft}.) Note that nonzero $\Delta$ adds interactions and also modifies the bandwidth; at $\Delta = 1$, the spinon velocity is given by $v_s = \frac{\pi}{2}J$ as determined by the exact Bethe ansatz result for this system \cite{descloizeaux1962}.

While clean zigzag chains show a gapless Raman response (\eqnref{eqn:cleanmf}), we find that when there are multiple zigzag domains, and hence multiple domains of $R_A$ (\eqnref{eqn:ra}), the Raman response becomes gapped. The gap occurs at a frequency $\omega_c$ where  $\omega_c = v_s (\pi/L_d)$ where $L_d$ is the size of the domain between the two domain walls (unit lattice spacing) and $v_s$ is the spinon velocity. Here we consider domains of a particular size; the generalization to a distribution of domain sizes is discussed below. The presence of the gap in the 2-particle continuum may be surprising since in computing the scattering spectrum, neither the Hamiltonian nor the ground state are altered. Indeed, the only difference between these responses is the presence of zigzag domain walls. 

In applied field, we find another surprising result: at a small finite fields the gap closes then reopens (Fig.\ \ref{fig:mf-bos-responses}, top row). With $L_d$ the domain size, the gap in the Raman spectrum closes at a small applied field of $h_c = \frac12 v_s (\pi/L_d)$ and reopens for $h^z > h_c$. We again emphasize that in this case, the Hamiltonian is blind to the presence of the zigzag domain walls. Moreover the domain walls are non-magnetic  defects, and naively one would not expect such a drastically different response in the presence of applied magnetic field.  

\subsubsection{Bosonization} 
        
In the clean zigzag chain, we previously stated that the mean field spectrum at the 2-particle level did not even qualitatively capture the delta function response computed using bosonization. This inconsistency begs the question as to whether these results of a singular magnetic field response domain walls are robust beyond mean field. We find that they are.

\begin{figure*}
    \centering
    \includegraphics[width=0.49\textwidth]{figures/rd-zigzag.png}
    \includegraphics[width=0.49\textwidth]{figures/rd3-zigzag.png}
    \includegraphics[width=0.49\textwidth]{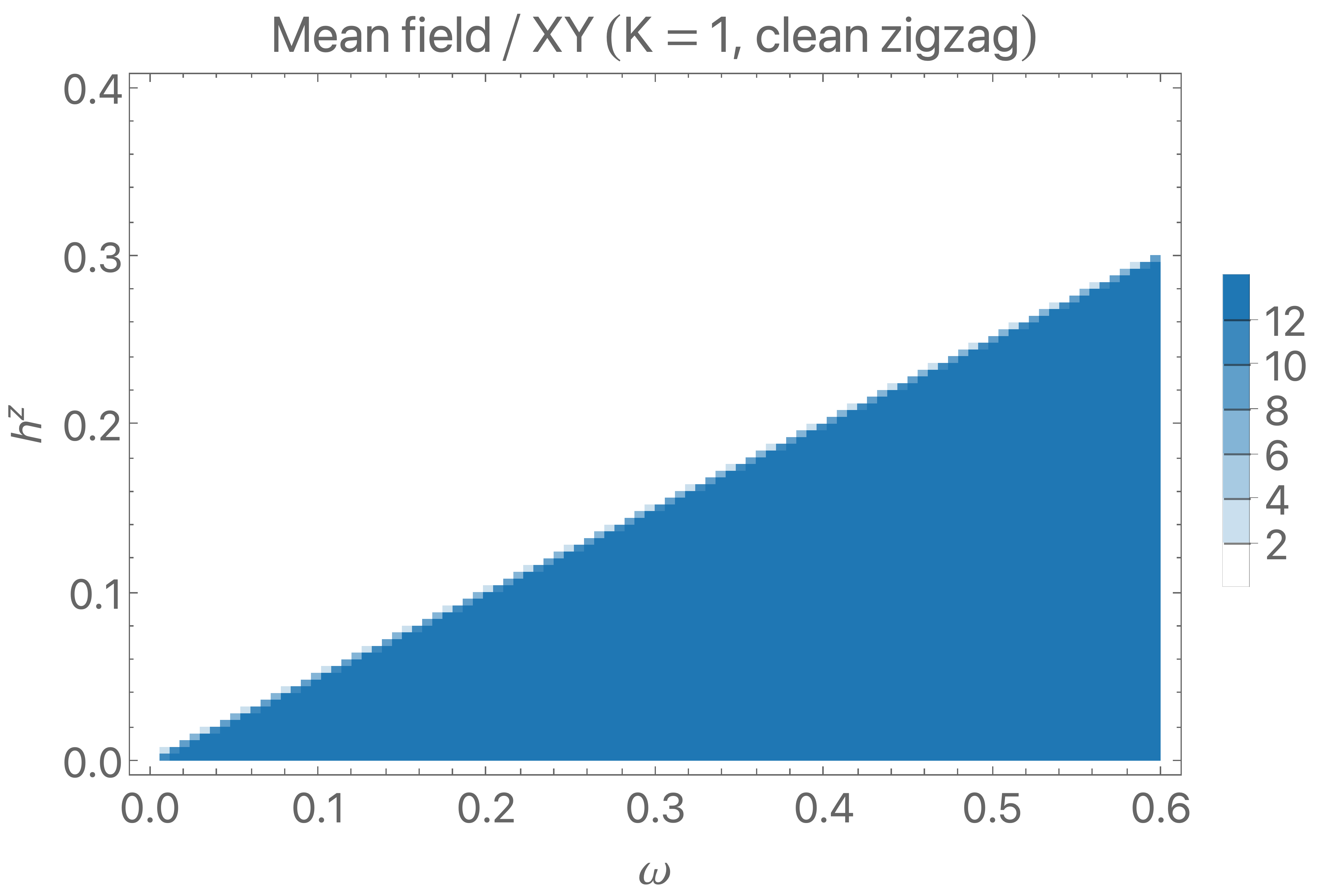}
    \includegraphics[width=0.49\textwidth]{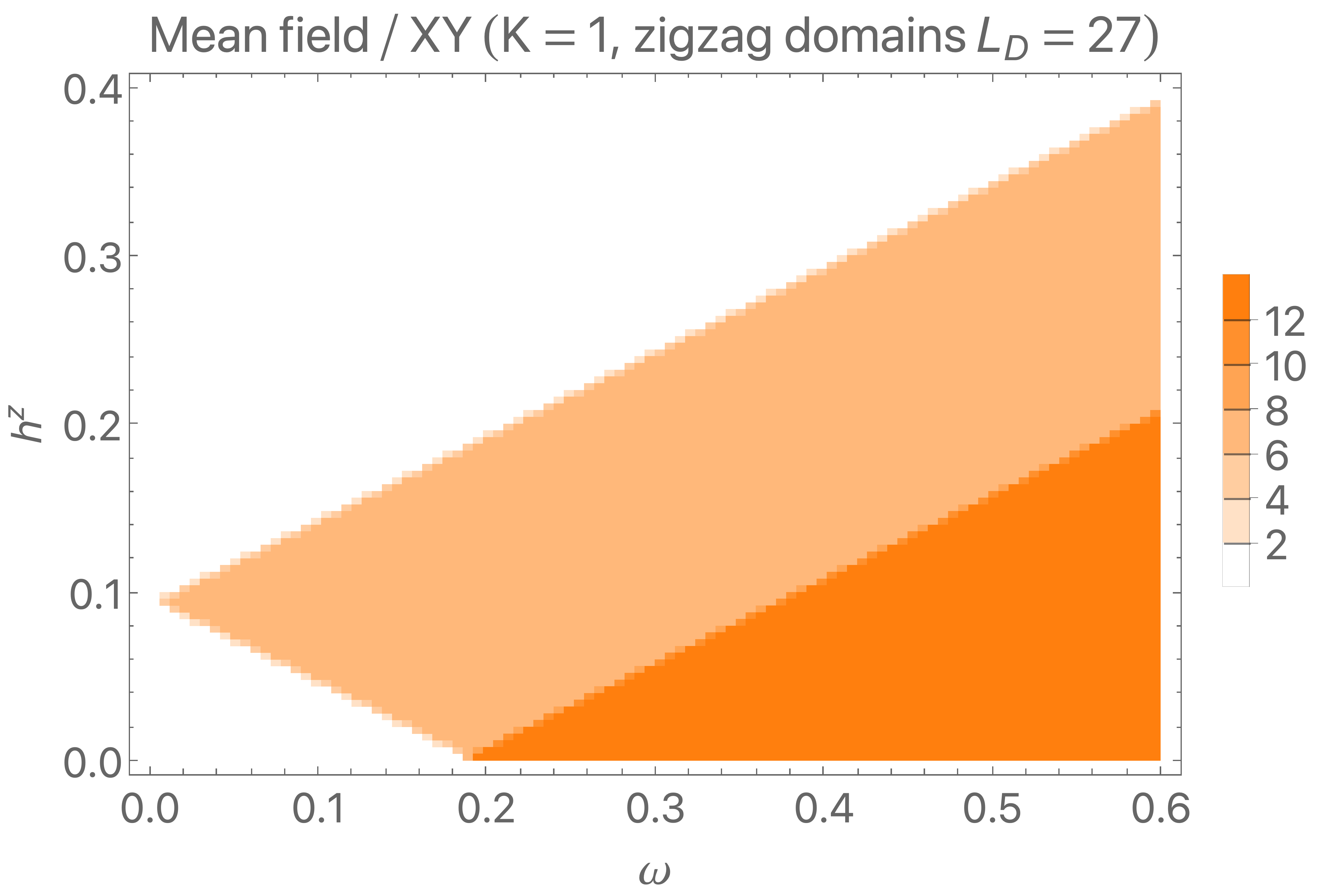}
    \includegraphics[width=0.49\textwidth]{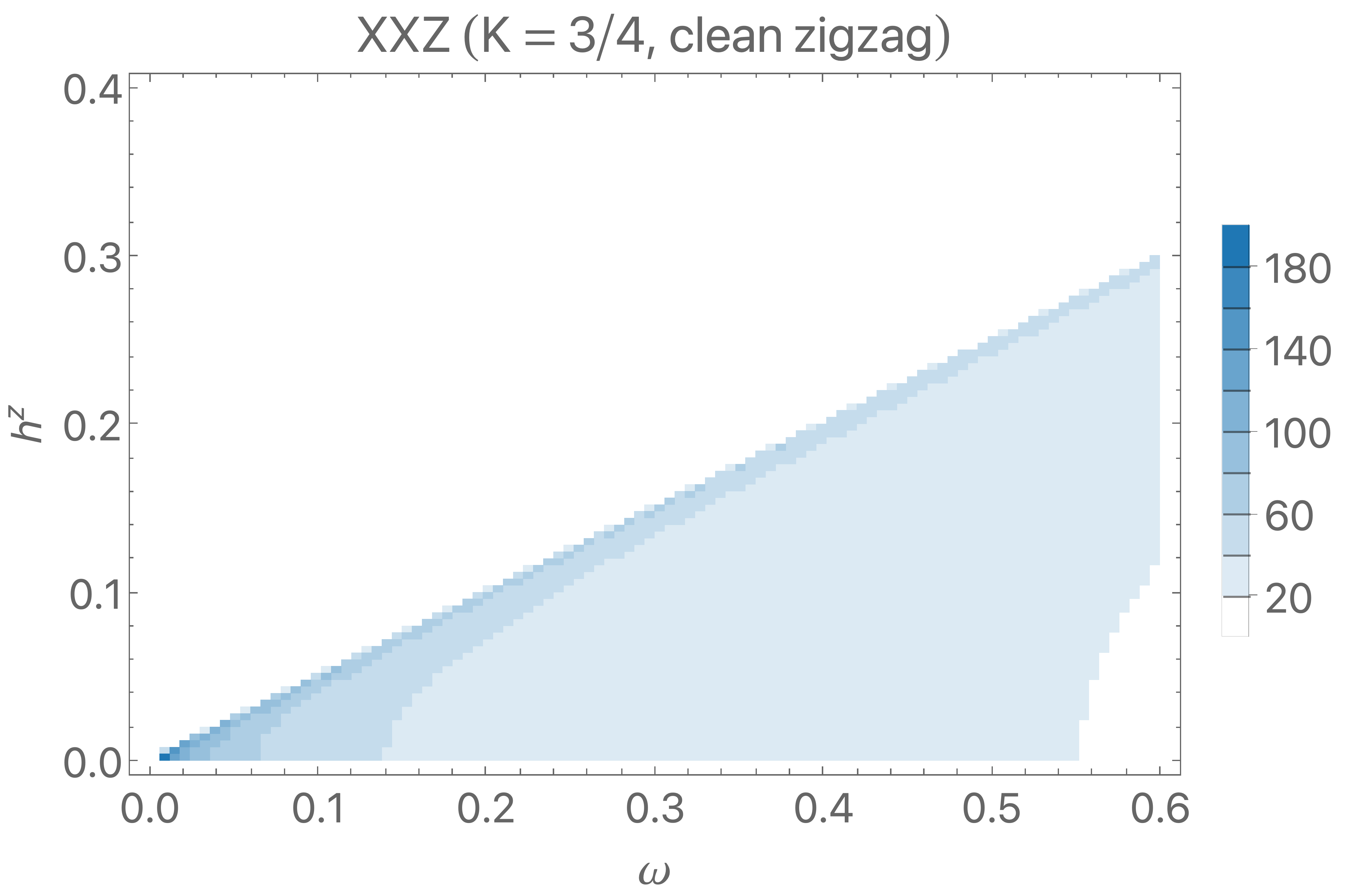}
    \includegraphics[width=0.49\textwidth]{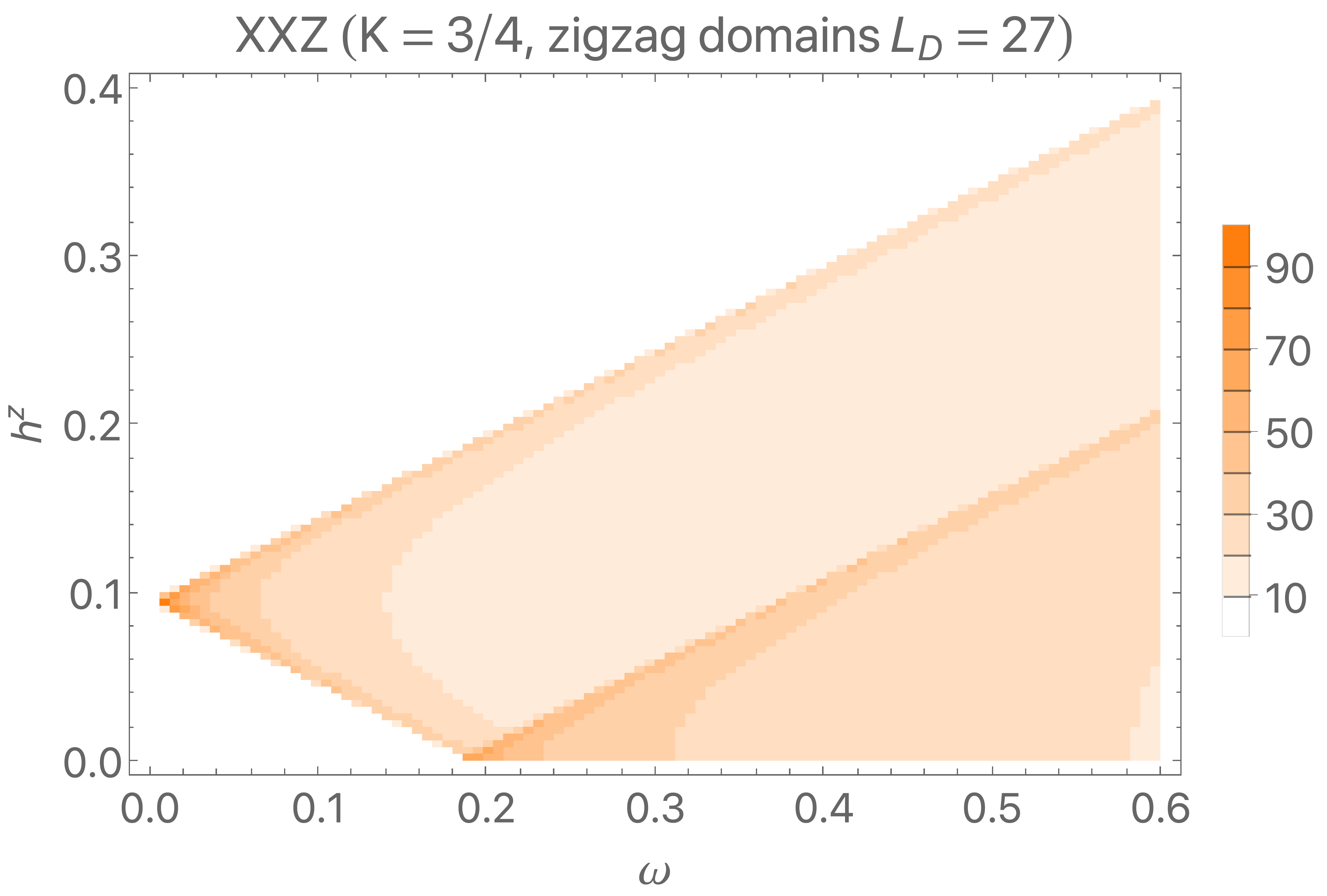}
    \includegraphics[width=0.49\textwidth]{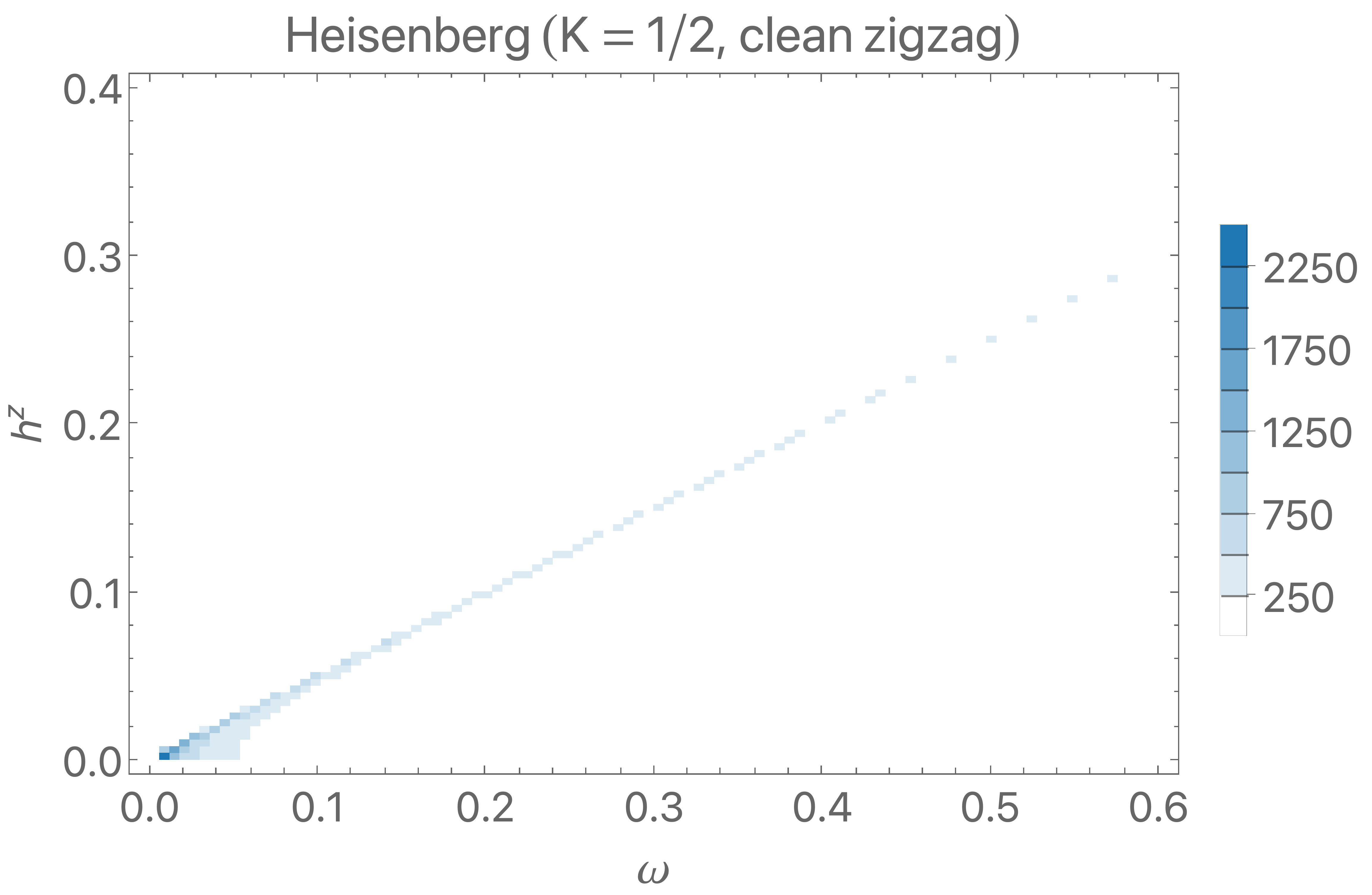}
    \includegraphics[width=0.49\textwidth]{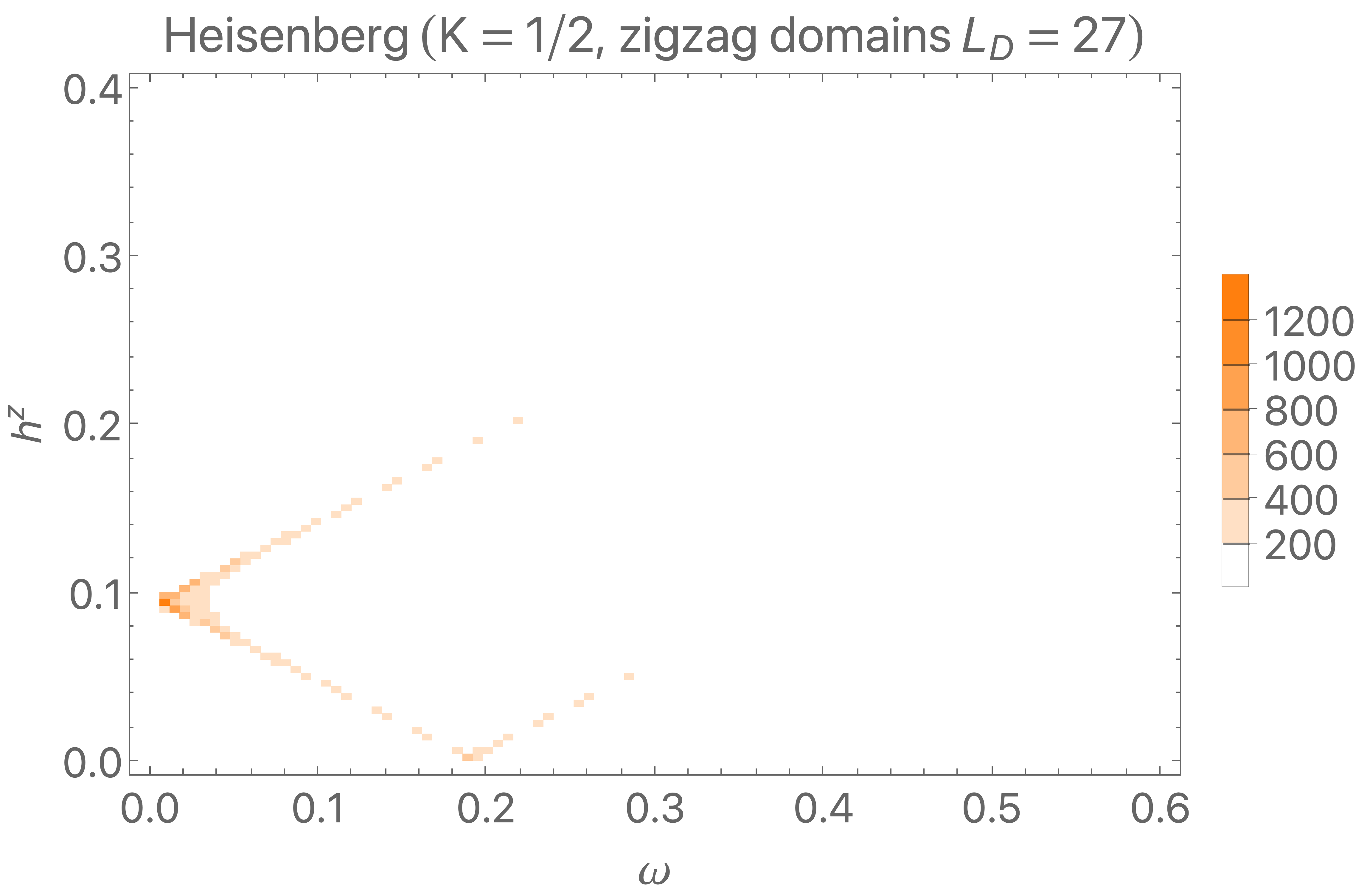}
    \caption{
    Low frequency magneto-Raman spectra computed via field theory bosonization (\eqnref{eqn:boschi}), for the clean zigzag chain $R_A$ (left column, blue) versus the $R_{DW}$ chain with multiple zigzag domains (of size $L_d{=}27$) (right column, orange). For ease of comparison, spinon velocity is fixed as $v_s = \pi/2$. Top row: XY model, $\Delta=0$, where mean field is exact (Luttinger parameter $K = 1$). Middle: a representative XXZ response for intermediate $\Delta$ (at $K = 3/4$). Bottom: The  $\Delta=1$ SU(2) Heisenberg Hamiltonian ($K = 1/2$), with a delta function singularity. Across the interaction range, the clean zigzag spectra shows gapless excitations whose gap increases linearly with applied magnetic field. In the presence of zigzag domain walls, a Raman gap is opened, but closes and then reopens with applied magnetic field.}
    \label{fig:mf-bos-responses}
\end{figure*}

At low frequencies the Raman response can be described analytically via bosonization of the interacting spinon model. (For details such as discussion of renormalization of Luttinger parameter $K$ with magnetic field, see  Appendix \ref{sec:appbos}.) At the 2-particle level for $R$ (though Hamiltonian interactions are captured by $K$) we use the density-density response function computed in bosonization. Linearizing momenta about the Fermi points of the spinon liquid, we find the vertex contribution $V_{kq}$ is constant. Hence the 2-particle Raman response of $R_q$ reduces to density-density correlation function within bosonization. At a temperature $T$ (setting $k_B=1$) this correlation function is a textbook computation and is given by \cite{giamarchi2004} 
\begin{align}\label{eqn:boschi}
 &   \chi_{\rho\rho}(q,\omega) \propto -\frac{\sin(\pi K)}{v_s}\left( \frac{2\pi T}{ v_s} \right)^{2K - 2}F(q, \omega) 
    \\
&F(q, \omega) = \prod_{\eta=\pm 1}
B\left(-i\frac{\omega + \eta v_s q}{4\pi T} + \frac{K}{2}, 1-K\right)
\end{align}
with $B(x,y)$ Euler's beta function $B(x,y)=\Gamma(x)\Gamma(y)/\Gamma(x+y)$. $K$ and $v_s$ are Luttinger parameter and velocity. The Raman response of $R_q$ is then $-\text{Im}(\chi_{\rho\rho}(q,\omega))$. The results for $T\rightarrow 0$, as a function of magnetic field (discussed further in Appendix \ref{sec:appbos} and Eqn.\ref{eqn:chibosh}) are plotted in Fig.\ \ref{fig:mf-bos-responses}. 

\begin{figure}[t]
    \centering
    \includegraphics[width=0.48\textwidth]{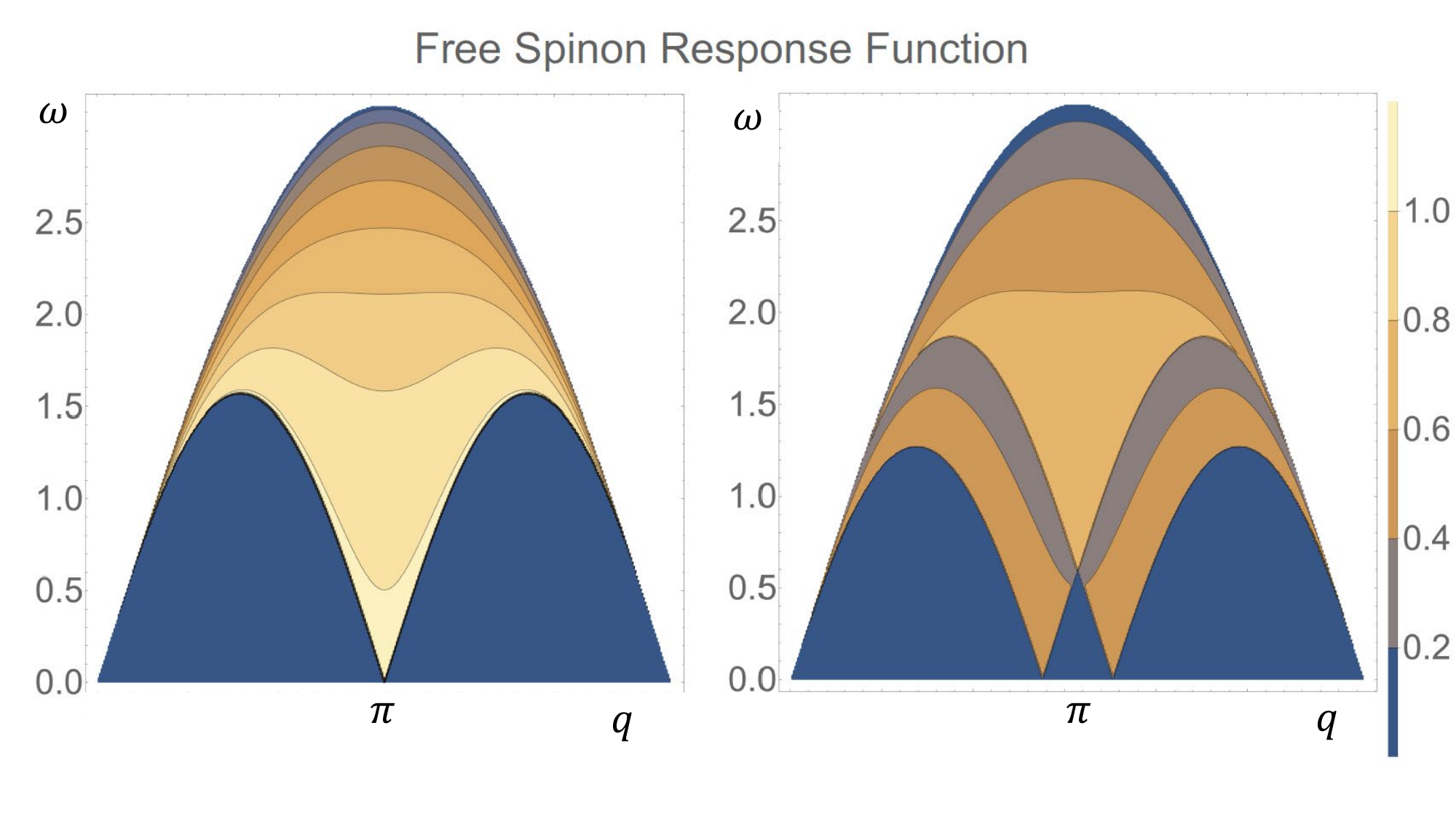}
    \caption{$\chi(q, \omega)$  two-fermion response function for a mean field free spinon Hamiltonian  with $v_s{=}\pi/2$, at zero field (left) and small applied field $h^z = 0.3$ (right). At this level the Raman spectrum is a weighted sum over $\chi(q)$ given by \eqnref{eqn:ifromchi}. For a clean zigzag chain, it is just a cut through $\chi$ at $q = \pi$. In the presence of zigzag domain walls, the Raman spectrum contains cuts at Fourier modes of the domain profile, $q \neq \pi$; for domains of typical size $L_d$, the most significant contributions come from $q = \pi \pm \pi/L_d$.}
    \label{fig:mf-free-spinons}
\end{figure}

\subsection{$\textit{O}(1)$ additive changes via $\textit{O}(\epsilon)$ wavevector shifts}\label{sec:explanation} 
In the previous sections, we described our results that non-magnetic crystal defects (zigzag domain walls) in a spin chain give a singular magnetic field response as measured by inelastic Raman scattering. This effect is not only captured in mean field, but it is also reproducible numerically (via TEBD) and analytically (at low energies using bosonization) in the presence of strong interactions. 
In this section, we describe the mean field (and associated setup for bosonization) analysis in detail, to explain how non-magnetic crystal defects induce a singular magnetic field response. These computational details will anchor the discussions in the following section.

We capture the effects of defects in the Raman operator by considering the Raman response of a sum of Heisenberg terms $R_j$ with coefficients $g_j$:
\begin{equation}\label{eqn:rq}
   R = \sum_j g_j R_j, \quad  R_j \equiv \mathbf{S}_j \cdot \mathbf{S}_{j+1}
\end{equation}
where the couplings $g_j$ are specified by the photon polarization factor of equation \eqnref{eqn:fleuryoperator}. We will call $g_j$ the bond profile of the Raman operator. 
For simplicity of notation here we restrict to Heinsenberg terms in $R$: XXZ anisotropy in $R$ is discussed in Appendix~\ref{sec:spinonmft}. 

By Fourier transforming $g_j$, the operator $R$ can be rewritten as a weighted sum over its Fourier modes as
\begin{equation}\label{eqn:genericr}
    R = \sum_q \tilde{g}_q R_q, \quad R_q = \sum_j e^{iqj} (\mathbf{S}_j \cdot \mathbf{S}_{j+1})
\end{equation}
where $\tilde{g}_q$ are the Fourier modes of $g$. The Raman response of a generic Raman operator $R$ is then given by 
\begin{equation}
    I(\omega) = \sum_{qq'} \tilde{g}_{q} \tilde{g}_{q'} \int dt \ e^{i\omega t} \braket{R_q(t) R_{q'}(0)}_0
\end{equation}

We now compute the correlation function $\braket{R_q(t) R_{q'}(0)}_0$. This object is only nonzero for $q+q' = 0$, and so it suffices to compute $\braket{R_q(t) R_{-q}(0)}_0$. We then define the Fourier transform of this correlation function to be 
\begin{equation}\label{eqn:chi}
\chi(q, \omega) = \int dt \ e^{i\omega t} \braket{R_q(t) R_{-q}(0)}_0    
\end{equation}
so that the Raman intensity of $R$ is given by
\begin{equation}\label{eqn:ifromchi}
    I(\omega) = \sum_{q} |\tilde{g}_q|^2  \chi(q, \omega)
\end{equation}
Raman spectra are a weighted sum over finite $q$ probes depending on the functional form of $g$. Even when photons carry effectively zero momentum, Raman is not always a $q = 0$ probe.

The response $\chi$ is easily understood within mean field. We use the Jordan-Wigner transformation to map spin operators to spinless fermionic operators and compute the response $\chi$ to lowest nonvanishing order in fermionic operators (Appendix \ref{sec:spinonmft}). At this order, $\chi$ is a   response function involving 4 fermion operators. This approximation becomes exact when $R$ Heinsenberg terms are replaced by XY terms, leading to $\chi^{MF}(q,\omega)$ given by
\begin{equation}\label{eqn:mf2particle}
    \chi^{MF} = \sum_{k_0}\frac{8}{2\pi} \frac{\sqrt{(2v_s\sin(q/2))^2 - \omega^2}}{(2v_s\sin(q/2))^2} f(k_0)(1-f(k_0-q))
\end{equation}
where $f(k)$ is the Fermi function evaluated at the energy $\epsilon_k$, with $\epsilon_k = -v_s\cos k$, and $k_0$ are the wavevectors that satisfy energy conservation $\omega + \epsilon_{k_0} - \epsilon_{k_0-q} = 0$. At low frequencies the relevant part of $\chi$ is just this $2k_F$ response with  a Heaviside step function $\Theta$,
\begin{equation}
    \chi(q, \omega) \sim  \sum_\pm \Theta(\omega - v|q\pm 2k_F|) 
\end{equation}   
We plot $\chi^{MF}$ in Fig.\ \ref{fig:mf-free-spinons}.

\begin{figure}[t]
    \centering
    \includegraphics[width=0.99\columnwidth]{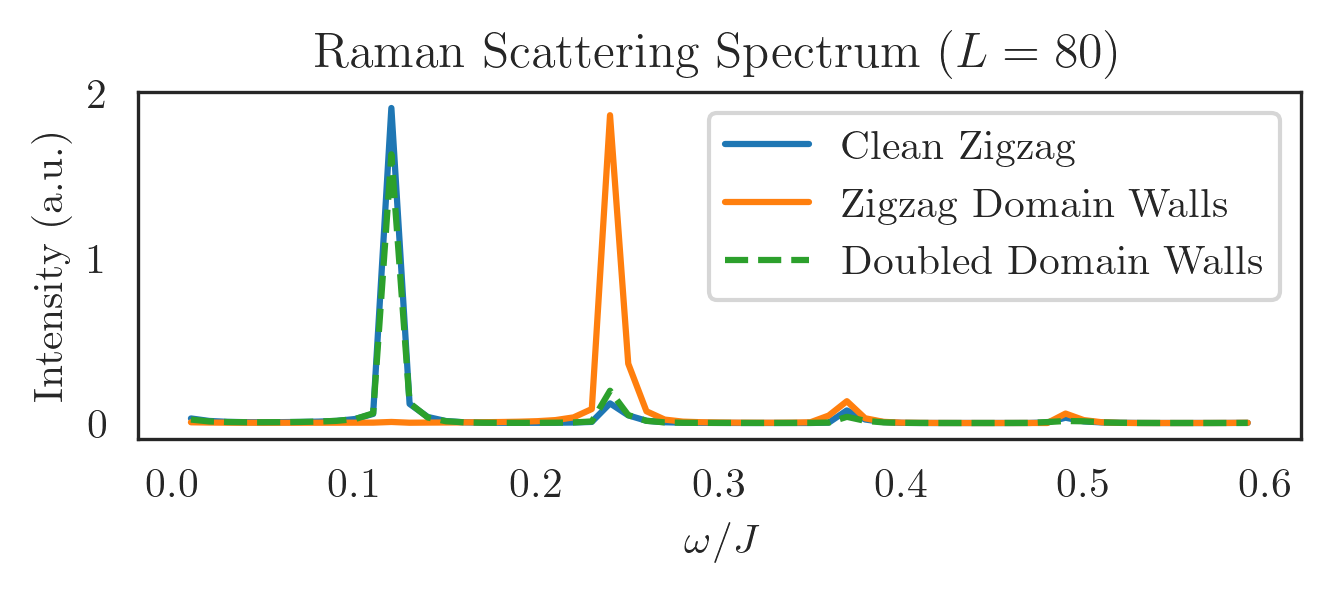}
    \caption{Inelastic Raman spectra computed with DMRG and TEBD for a finite size $L{=}80$  Heisenberg spin-1/2 chain with zero domain walls (solid, blue), two domain walls (solid, orange), and doubled domain walls (dashed, green). The two domain wall system is the same as shown in the right-hand column of Fig.\ \ref{fig:tebd}. The doubled domain wall system has two domain walls on sites $j_1{=}39$ and $j_2{=}41$. The three systems have the same Hamiltonian, and only vary in the choice of Raman operator. For doubled domain walls, the zero magnetic field gap associated with the presence of isolated domain walls is no longer observed, showing that the effect of domain walls arises from their $Z_2$ character as topological defects, rather than from local defect effects.}
    \label{fig:z2-raman-cut}
\end{figure}

Raman scattering of a clean zigzag spin chain with zero domain walls is  a $q = \pi$ probe. Here $R_A \propto \sum_j (-1)^j \textbf{S}_j \cdot \textbf{S}_{j+1}$ with bond profile $g_j = (-1)^j$, giving a Fourier transform of $\tilde{g}_q \propto \delta_{q,\pi}$. Thus we use \eqnref{eqn:mf2particle} at zero magnetic field and at $q = 2k_F = \pi$.  At low frequencies, $k_0 = \pm k_F$ where the Fermi wavevector is $k_F=\pi/2$ at zero magnetic field.  We find the gapless response $I(\omega) \propto \sqrt{1 - (\omega/2v_s)^2}$.

Zigzag domain walls create a zero-field gapped response because the Raman scattering probe for the spin chain is shifted away from $q = \pi$. The amounts of the shift across various wavevectors is given by the Fourier transform power spectrum of the bond profile; if domains have a typical size $L_d$, this Fourier transform will be sharply peaked at $\pi\pm \pi/L_d$. For large $L_d$, the zero-field gap is $\omega_c \approx v_s \delta q = v_s(\pi/L_d)$. This gap closes and then reopens in applied magnetic field because the Fermi momentum $k_F$ of the fermionic spinons changes in applied field. The magnetic field is the chemical potential of  spinons. In the presence of zigzag domain walls, the Raman spectrum as probing excitations at $q$ away from $\pi$. In applied field, however, $k_F$ also shifts from $\pi$ such that the finite momentum probed by Raman scattering can be in resonance with the new Fermi momentum (i.e. $q \pm \delta q = 2k_F$). Hence the gap closes in applied fields. At larger magnetic fields $2k_F$ shifts further and the gap reopens.

Creating a small  density $\epsilon$  of zigzag domain wall defects involves the modification of an order-1 fraction of the system, i.e.\ flipping half of the domains. This feature of topological defects enables them to modify the Raman response by an order-1 amount, ie the difference  between the Raman response of the $\epsilon$-density domain wall system and the clean system is order-1, $R(\epsilon) - R(0) \sim 1$. However the region in frequency where this difference occurs is small, extending over a frequency window of order $\epsilon$, associated with the order $\epsilon$ wavevector shifts created by the domain walls.

\section{Generalizations and Discussion}\label{sec:discussion} 

\subsection{$Z_2$ Characteristic and other distinctions between topological defects and local defects}

Domain walls  are topological defects in that to create them one must create an entire domain. This nonlocality enables them to produce an $\textit{O}(1)$ change in the response theory. Associated with that nonlocality, domain walls also have a $Z_2$ character. This $Z_2$ character is indeed seen in the response to such defects: when the spatial separation between domain walls becomes small, the anomalous magnetic field response is no longer observed in our numerics (Fig.\ \ref{fig:z2-raman-cut}). In contrast, local (non-topological, or, geometric) defects don't show a $Z_2$ character and only add a small additive component to the response. We now discuss these two distinctions in detail. 

Isolated domain wall defects shift the probed wavevector. For a single domain, the wavevector probed in the Raman response is $q = \pi$. The presence of a topological defect in the form of a zigzag domain wall causes the Raman response to shift from $q = \pi$ to $q = \pi + \delta q$. The length of zigzag domains $L_d$ fixes the small wavevector $\delta q = \pi/L_d$. Put another way, the Raman operator for the system changes from $R_A$ (or $R_{q = \pi}$) to $R_{\pi \pm \delta q}$. The subsequent Raman response changes from $\chi(q = \pi, \omega) \to \chi(q = \pi + \delta q, \omega)$. 

Doubled (nearby) domain walls are a local defect. They are locally creatable, and thus do not change the wavevector probed by the clean system, and instead only add a small amplitude component on top of the clean system response. In terms of wavevectors, the doubled domain wall is local in real space and hence spread widely in reciprocal space, mostly on large wavevectors $Q$ whose response $\chi(Q,\omega)$ vanishes at low frequencies. The crossover from the topological nature of well separated domain walls to local nature of two domain walls in close proximity to each other can also be seen numerically (Fig.\ \ref{fig:z2-raman-cut}). 

Another example of local (non-topological) defects is a small amplitude variation in the zigzag angle $\theta_0$ for the spin chain. It does not shift the wavevector probed in Raman scattering for the clean system. Suppose $\theta_j$ is the bond angle on the $j$th bond, and we let $\theta_j = \theta_0 + \epsilon f_j$ for a mean angle of $\theta_0$, a small amplitude $\epsilon$, and an arbitrary function of position $f_j$. In this case, we have (\eqnref{eqn:geoepsilon})
\begin{equation}
    I(\omega) = A \chi(\pi, \omega) + \epsilon^2 \cos(2\theta_0)\sum_q |\tilde{f}_q| \chi(q+\pi, \omega)
\end{equation}
where $A$ is $\textit{O}(1)$. Here, the Raman response is composed of the clean system response at $q = \pi$ and a small $\textit{O}(\epsilon^2)$ component. Whereas topological defects shift the Raman response to $q \neq \pi$, geometric defects only add a small component rather than shifting the wavevector probed.

\subsection{Role of Average Symmetry}
We note that the disordered system with domain walls still has an average symmetry which is key to the observations described above. On average, neither domain is preferred over the other. This average symmetry ensures that the domains occur with equal probability, and hence that the $q=0$ component of the domain structure Fourier transform vanishes exactly. This ensures that the distribution of $\delta q$ has no weight at $q=0$ and hence no gapless response remains in the Raman response at zero field. 

A related feature is that the average symmetry ensures exact cancellation of the contribution of each single domain. Writing the Raman operator $R$ as a sum over domain $D_1$ and domain $D_2$ terms $R_{1}, R_{2}$, the dynamical $R$ correlation function clearly contains a direct term of $R_{1} R_{1}$ and $R_{2} R_{2}$, which at zero field produce a gapless response. What happens to that gapless response? The answer is that it is exactly cancelled out by the cross terms $R_{1} R_{2}$. This exact cancellation requires the average symmetry.
    
\subsection{Effect of Zigzag Domain Walls as local defects within the Hamiltonian} 
Above we consider the effects of domain walls purely in modifying the domains in the Raman operator, in order to isolate the effects of crystalline topological defects within a response theory. However in addition to this effect, the domain wall does create a local defect both in $R$ and in $H$. We now discuss these local effects.  In the particular 1D model considered here, perturbations to the Hamiltonian are easily RG-relevant due to the 1D physics, so we will also discuss the resulting effects in a 1D chain, noting however that local defects will not have such strong effects in 2D or 3D systems. 

Microscopically it is clear that local defect effects arise at domain walls.  Modifications in the Hamiltonian would arise from changes in the spin exchange coupling due to any changes in the bond length. The relative orientation of these bonds with photon polarizations would also affect their bond strength in the Raman operator.

In 1D, local bond perturbations to the Hamiltonian are relevant in the renormalization group (RG) sense \cite{eggert1992}. The fixed point of the RG flow breaks apart the system into two open chains. For a weak bond, the end points of the chain are precisely the two adjacent lattice sites between which the exchange coupling is reduced; for a strong bond, a singlet is formed, but the chain is cut away from this singlet. 

When zigzag domain walls are present in spin chains and modify the Hamiltonian locally, the RG fixed point of the system is a set of fragmented finite size chains of varying length given by the size of each zigzag domain.
 Finite chains of length $L_d$ have a true finite size gap, hence a gapped Raman response, associated with the finite size wavevector $q = \pi \pm \pi/L_d$. Magnetic fields tune the chemical potential through these finite sized wavevectors. Thus in both the limit where the Hamiltonian breaks apart into fragmented chains and in the limit where it does not (and e.g. is unmodified), the presence of zigzag domain walls gives a Raman response at a finite wavevector away from the wavevector probed by the clean system.

Above we have also computed the Raman spectra numerically in finite size systems. This corresponds to an intermediate case, where the Hamiltonian perturbation is included, but the fixed point of fragmented chains is not yet reached. This can occur physically if the RG flow is arrested by some effective finite size effects, e.g. a moderately weak bond from a domain wall where the next domain walls have already completed an RG flow to decoupling, such that the domain wall occurs within a finite chain fragment. Numerically for $L=80$ sites, the Raman spectra are quite similar to those of an unperturbed Hamiltonian. We also look at varying the strength of the horizontal domain wall bond within $R$, which can vary corresponding to the strength of the corresponding bond in $H$, in a manner that also depends on photon polarizations. The results are again qualitatively independent of this modification (see Fig.~\ref{fig:realcolorplots}).

We next turn to comparing the two distributions of wavevectors in both the unperturbed and perturbed $H$ limits.     

\begin{figure}[t]
    \centering
    \includegraphics[width=0.48\textwidth]{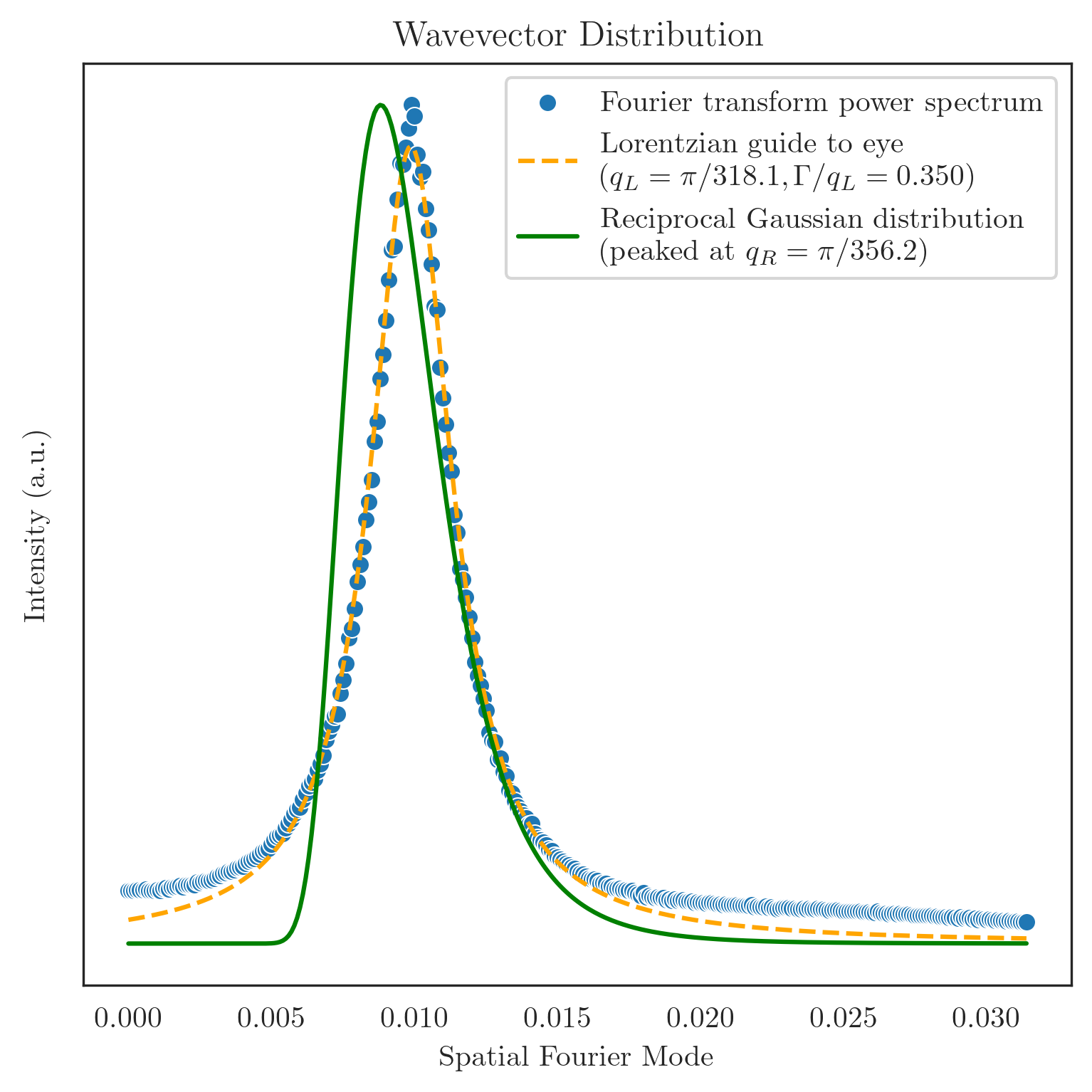}
    \caption{Distributions of wavevector deviations from $\pi$ that are relevant for zigzag domains whose sizes follow a Gaussian distribution with mean $\mu = 300$  and standard deviation $\sigma = 100$. Blue circles: numerically averaged power spectrum of Raman operator bond profile ($10^3$ chains of length $10^5$), giving the wavevector distribution in the unmodified Hamiltonian limit. Orange dashed: Lorentzian centered at $q_L$ with FWHM $\Gamma$ as a guide to the eye. Green solid: reciprocal Gaussian distribution, giving finite size gaps in the fragmented chain segments limit.}
    \label{fig:disorderavgft}
\end{figure}
    
\subsection{Distribution of Domain Sizes}
The above sections considered the case of a particular size of zigzag domains. Here we consider the realistic case of a distribution of domain sizes. The crystalline domain size is expected to be a random distribution with a typical length scale set by the competition between the cost of domain walls and the preference to have a single domain. For example, this can be modeled in an equilibrium approximation by considering a random ``field'' that locally prefers one of the two crystalline domains. The resulting Imry-Ma mechanism  \cite{imry1975} for discrete order in 1D gives domains at the Larkin length. Indeed generally for disordered elastic media subjected to random fields (such as a local preference for one crystalline domain), the disorder seen by domain walls has long range correlations even when the microscopic disorder is short ranged correlated. 

\begin{figure}[t]
    \centering
    \includegraphics[width=0.48\textwidth]{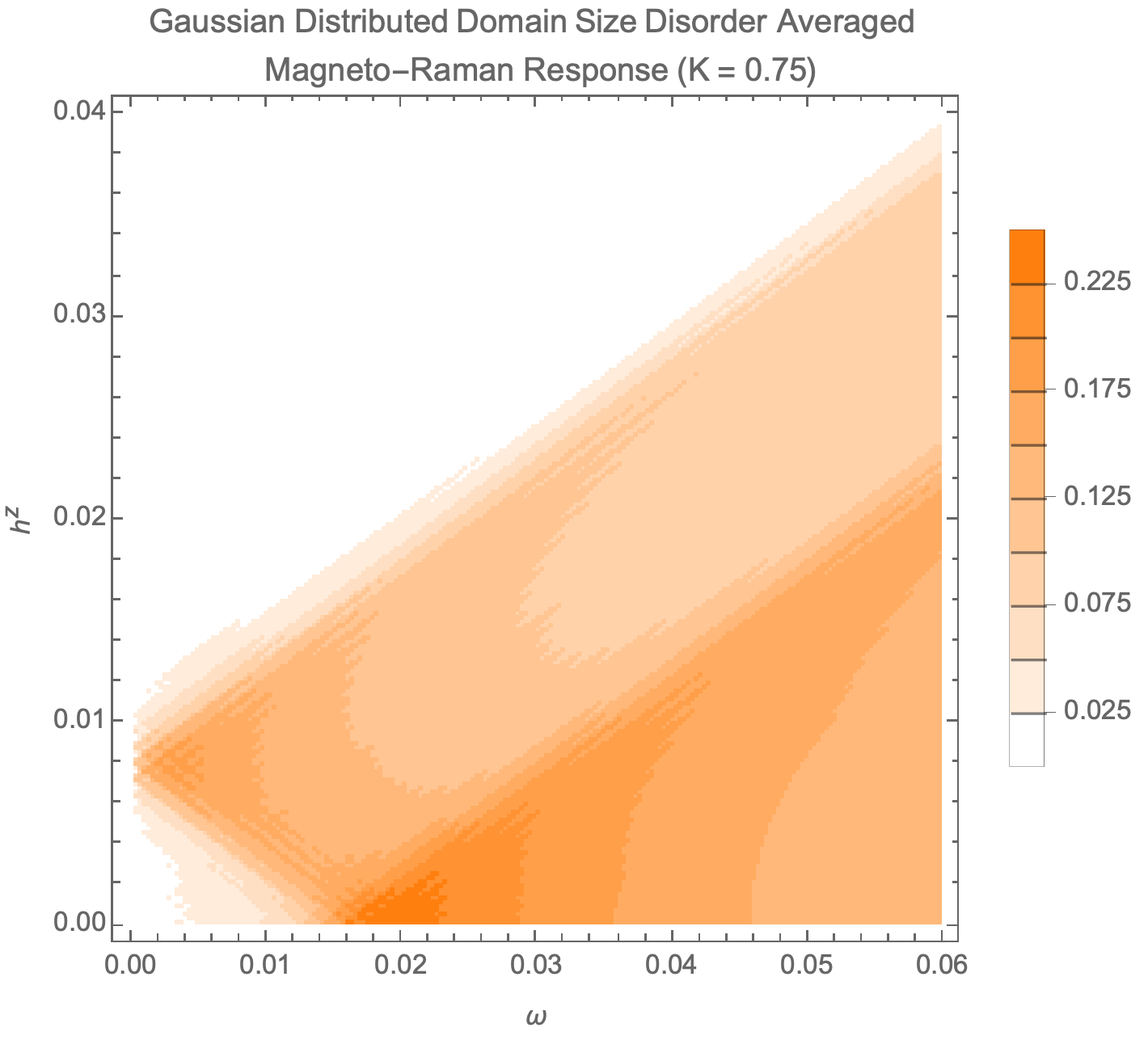}
    \caption{Magneto-Raman spectrum for zigzag domains of $R_{DW}$ with a distribution as in Fig.~\ref{fig:disorderavgft}, with Luttinger parameter $K = 3/4$ and spinon velocity $v_s = \pi/2$. }
    \label{fig:hwgaussian}
\end{figure}

To model a typical length scale arising from various equilibrium and nonequilibrium mechanisms, we consider a Gaussian distribution of domain lengths with mean $\mu$ and variance $\sigma^2$. In the limit where domain walls don't substantially modify the Hamiltonian, the resulting Raman spectra are set by integrating over wavevectors weighted by the Fourier transform power spectrum of the bond profile of the Raman operator (\eqnref{eqn:ftpoisson}). Numerically computing the power spectrum we find  it to be approximately a Lorentzian centered at $q_L \approx \pi/\mu$. In the fixed point limit where the Hamiltonian  is strongly modified by the domain walls, the system becomes an ensemble of fragmented spin chains each with their own finite size gaps. These finite size gaps follow a reciprocal Gaussian distribution, peaked at $q_R = \pi (-\mu + \sqrt{\mu^2 + 8\sigma^2})/(4\sigma^2)$. As a general statement for the mean of the reciprocal distribution independent of the $L_d$ distribution, note that the wavevector shift is $dq = \pi/L_d$  for each chain fragment of size $L_d$, so its expectation value is bounded from below by $\pi/\overline{L}_d$ by Jensen's inequality: $\overline{dq} > \pi/\overline{L}_d$, as indeed seen for the reciprocal Gaussian. Both distributions (Fig.~\ref{fig:disorderavgft}) are sharply peaked, resulting in magneto-Raman spectra with a soft gap but still showing the features described above (Fig.~\ref{fig:hwgaussian}).

The importance of the long range effects that set a typical domain size can be seen by considering an opposite limit, where every site has an independent probability of hosting a domain wall. The resulting distribution of domain sizes is Poisson. This is  unphysical: crystalline domains would not form a Poisson distribution, since for example it would be energetically preferred to shift two nearby domain walls together and annihilate them at the small cost of flipping an unfavored domain. A Poisson distribution would however be expected for locations of non-topological local defects with no associated domains. Such a Poisson distribution  (``random telegraph signal'') would produce a $q=0$ centered Lorentzian power spectrum. It would show a gapless Raman signal at nonzero fields, but would not show the singular gap behavior of domains with typical length scale.

\subsection{Role of Spinon Liquid State} 
The singular magnetic field responses of zigzag domain walls described above were computed within the Luttinger Liquid ground state of $H_0$. Here we show that they rely on the special properties of this gapless spinon phase, and in particular that they are absent in systems with ferromagnetic order, antiferromagnetic order, and in the gapped AKLT type phases allowed for integer spin chains.

For AKLT states, the low energy responses cannot occur simply because the gap is too large. For example the gap is $\Delta \approx 0.41 J$ for $S = 1$ \cite{affleck2005}. In magnetic fields this gap closes only for $h > \Delta$ at which the system experiences magnon condensation \cite{affleck1991}. Since the AKLT chain remains gapped for fields below $\Delta$, the effects above, for small magnetic fields and small frequencies, can not occur. 

To study the magnetically ordered phases we compute the response of the operator $R_q$ from equation \eqnref{eqn:rq} within linear spin-wave theory (LSWT), to compute a 4-magnon response function analogous to equation \eqnref{eqn:chi}. It is well known \cite{mermin1966} that quantum fluctuations in 1D diverge and obstruct magnetic order. Nonetheless, we employ LSWT in order to gain intuition for how quasi-1D spin chains may behave in the presence of magnetic order mediated by interactions in 3D. 
The contrast with the spinon liquid is clear. 
Whereas an applied magnetic field shifts the Fermi momentum of a spinon Fermi surface, and may tune gapped excitations to become gapless, bosonic magnon excitations do not experience such an effect. Indeed, for 1D magnons we find no similar anomalous magnetic field response that is probed by Raman scattering. 

We first consider magnons in a ferromagnetic phase in applied magnetic field. 
In LSWT, the ground state is the magnon vacuum. The Raman operator $R_q$, however, is magnon density excitations at finite $q$. Explicitly we find $R_q = \sum_k \Gamma_{kq} a_k^\dagger a_{k+q}$ where the vertex $\Gamma_{kq} = S(e^{-i(k+q)} + e^{ik} - (1+e^{-iq}))$. Since $R_q$ annihilates the ground state, the $T = 0$ Raman scattering spectrum within linear spin wave theory vanishes at zero field. Although applied magnetic fields open a gap in the single particle spectrum, the classical ground state remains unchanged.

Next we consider the antiferromagnetic phase. In zero applied field the Raman response is a four magnon correlator. 
In  an applied field, the classical ground state of the antiferromagnet cants. Unlike the previous cases, where the Raman operator $R_q$ is a magnon density excitation, in the antiferromagnetic phase in an applied field the lowest order contribution to the Raman operator is a single magnon excitation (Appendix \ref{sec:afmlswt}). 
As such, the lowest order contribution to the Raman response $\chi^{AFM}(q, \omega)$ qualitatively follows the single magnon dispersion, with a $\pi$ shift in momentum, from the $\pi$ wavevector associated with the classical N{\'e}el ordered ground state from which spins cant in applied magnetic field. 
For nonzero domain wall densities, the Raman response measures especially $q = \pi \pm \delta q$ with $\delta q$ small. Antiferromagnetic magnons at $q = \pi$ are gapless, and in applied field they remain gapless. Excitations away from $q = \pi$ always remain gapped. Moreover, since magnons are bosonic and have no Fermi momentum, applied fields cannot close the gap opened by the presence of domain walls. Consequently, although domain walls open a gap in the spectrum, this gap would not close in the presence of an applied field.

\section{Outlook}

In this work we argued that in certain settings, crystalline topological defects can modify the response of an electronic system not just by changing the electronic state, but also by changing the response theory operator. In particular this arises in experiments that involve electric fields or photon scattering. We presented a proof-of-principle toy model using Raman scattering on a zigzag spin half chain with zigzag crystalline domain walls. In this toy model, even in the limit where domain walls only enter the Raman operator and not the Hamiltonian, still they produce singular effects in the Raman spectra, including singularities in applied magnetic fields. Such effects may be otherwise unexpected from nonmagnetic crystal disorder.

One way to understand the response is as arising from an effective shift of the wavevector probed by the Raman scattering (which is conventionally at $q=0$). Intriguingly, this wavevector shift arises from dimerization domain walls, which are here only within the Raman operator; in contrast dimerization domain walls in a spin-half Hamiltonian carry spin-half modes protected by the spin half quantum anomaly. 
Is there also an aspect of the Raman response here that is associated with a quantum anomaly? The Raman operator is not a Hamiltonian, so the framework needed to answer this question does not exist. However even though Raman is a scattering experiment, i.e.\ its theory is not a conventional linear or perturbative response theory, it could be considered to be a linear response theory with a contrived probe function which is the Raman operator. Nonlinear responses at zero frequency then involve this operator being added to the Hamiltonian. The resulting modified Hamiltonian would then carry explicit dimerization, and domain walls of this dimerization could carry anomaly-protected spin-half modes. Any such effects would necessarily be non perturbative. Could there be such quantum anomaly features associated with response theory operators? We leave this question for future work.

Other future avenues should explore the modification of response theories in higher spatial dimensions, and in particular as a way to modulate experimental response operators to tailor them as probes of particular otherwise-hard-to-probe quantum entangled phases. Already even simple Raman spectroscopy has been shown to be a useful probe of quantum spin liquid phases  \cite{perkins2008, knolle2014, perreault2015}. The ability to modify the effective wavevector probed, or make further modifications with real space resolution, just by adding crystalline topological defects, could be of much help. Importantly, crystalline topological defects may be a mild type of disorder for some quantum states and hence such defects can be added even intentionally, without destroying the desired quantum state. In this way the addition of crystalline topological defect can serve as a way to tune experimental response theory probes while preserving the quantum state.

\section*{Acknowledgements}
The authors acknowledge helpful discussions with  Ehud Altman, Erez Berg, Natalia Drichko, Zhigang Jiang,  Yuan-Ming Lu, Martin Mourigal, Tyrel McQueen, Colin Parker, Natasha Perkins, Daniel Podolsky, Hide Takagi, Achim Rosch, and Masaki Oshikawa. 
This work was performed in part at the Kavli Institute
for Theoretical Physics, which is supported by the National Science Foundation under Grants No.\ NSF PHY-1748958 and PHY-2309135.
This work was also performed in part at the Aspen Center for Physics, which is supported by National Science Foundation grant PHY-1607611.



\section*{Appendix}
\appendix

\section{Numerical Details}
Numerical simulations of Raman spectra were performed using the TeNPy library \cite{hauschild2018}. The ground state $\ket{\psi}$ is computed  using  DMRG (density-matrix renormalization group), initialized as a matrix-product state with N{\'e}el order. The Hamiltonian $H$ and Raman operator $R$ for a given simulation are written as matrix-product operators. To compute the Raman intensity, we compute the correlation function $\braket{\psi | R(t) R(0) | \psi}$ and Fourier transform to the frequency domain. 
Magnetic field values were taken in steps of $0.1$, and additional steps of $0.02$ at small fields (below 0.2 for the main text figures, below 0.1 for the appendix figures).

To compute $\braket{\psi | R(t) R(0) | \psi}$, we compute the equivalent quantity $e^{i\epsilon_0 t} \braket{\psi | R | \phi(t)}$, where $\epsilon_0$ is the ground-state energy, $\ket{\phi(0)} = R\ket{\psi}$, and $\ket{\phi(t)} = e^{-iHt}\ket{\phi} = e^{-iHt} R\ket{\psi}$. DMRG produces both $\ket{\psi}$ and $\epsilon_0$, and time-evolution of $\ket{\phi}$ is performed using time evolving block decimation (TEBD). 

The ground state of a finite size ($N = 80$) open $S = 1/2$ chain was found using DMRG with $10^{-10}$ precision in the ground state energy $\epsilon_0$. Bond dimension of 100 was found to be enough for convergence. Singular values were truncated below $10^{-10}$.

TEBD was performed using time step of \texttt{dt} = 0.0628 at Suzuki-Trotter order 4 with a built-in optimization described in \cite{barthel2020}. Numerical time evolution was performed for $10^5$ time steps. Bond dimension of 100 was again found to be enough for convergence. (recall we are computing time evolution with a local Hamiltonian starting from a ground state, so entanglement growth is relatively weak.) Singular values were truncated below $10^{-12}$. These parameters were chosen to obtain a frequency resolution $\Delta \omega \approx 0.01$ and reduce Trotterization error.

\section{Mean Field Theory and $R$ XXZ anisotropy}\label{sec:spinonmft}
In the main text we compute Raman responses using a mean field treatment. In this section for completeness we walk an interested reader through how to do the mean field computation. Within mean field, we take the  Hamiltonian to be a free spinon theory given by
\begin{equation}
    H = \sum_k \epsilon_k c_k^\dagger c_k, \quad \epsilon_k = -v_s \cos k
\end{equation}
with spinon operators obtained via a Jordan-Wigner transformation and, for Heisenberg Hamiltonian, the dispersion bandwidth set to the Bethe ansatz results for the spinon dispersion \cite{descloizeaux1962}. For the Raman operator $R_q = \sum_j e^{iqj} (\mathbf{S}_j \cdot \mathbf{S}_{j+1})$ we then seek to compute the dynamical correlation function $\braket{R_q(t) R_{q'}(0)}_0$ on the ground state. 

$R_q$ may generically contain XXZ anisotropy, presumably corresponding to the XXZ anisotropy of the Hamiltonian. Consider an XXZ operator 
\begin{equation}
R_q = \sum_j e^{iqj} (S_j^x S_{j+1}^x + S_j^y S_{j+1}^y + \alpha S_j^z S_{j+1}^z)
    V_{kq} V_{k-q,-q} 
\end{equation}
Heisenberg  couplings in the Raman operator corresponds to $\alpha = 1$. 
We first express the Raman operator $R_q$ in terms of Jordan-Wigner spinons. 
\begin{align}
    R_q &= \sum_j e^{iqj} \left( \frac12 \left( S_j^+ S_{j+1}^- + h.c. \right) + \alpha S_j^z S_{j+1}^z \right) \\
    &= \sum_j e^{iqj}\left( \frac12 (c_j^\dagger c_{j+1} + h.c.) + \alpha (n_j - \frac12)(n_{j+1}-\frac12 ) \right)
\end{align}
Next, approximate $R_q$ to quadratic order in spinon operators. 
\begin{align}
    R_q \rightarrow \frac12 \sum_j e^{iqj}\left( (c_j^\dagger c_{j+1} + h.c.) - \alpha(n_j + n_{j+1}) \right)
\end{align}
We then Fourier transform spinon operators with the convention $c_j = \frac{1}{\sqrt{N}}\sum_k e^{ikj} c_k$. To lowest order in fermionic operators, the inelastic part of the Raman operator $R_q$ in terms of Jordan-Wigner spinons is given by
\begin{equation}
    R_q = \sum_k V_{kq} c_k^\dagger c_{k-q}
\end{equation}
with the vertex given by
\begin{equation}
    V_{kq} = \frac12 (e^{i(k-q)} + e^{-ik} - \alpha(1 + e^{-iq}))
\end{equation}

The Raman operator $R_q$ is a spinon density excitation at momentum $q$. At time $t$, the operator $R_q$ is given by
\begin{equation}
    R_q(t) = \sum_k e^{i(\epsilon_k - \epsilon_{k-q})t} V_{kq} c_k^\dagger c_{k-q}
\end{equation}
The desired dynamical correlation function $\braket{R_q(t) R_{q'}(0)}_0$ can then be written as
\begin{equation}
    \sum_{kk'} e^{i(\epsilon_k - \epsilon_{k-q})t} V_{kq} V_{k'q'} \braket{(c_k^\dagger c_{k-q})(c_{k'}^\dagger c_{k'-q'})}_0
\end{equation}
where the first two fermionic operators are time ordered before the latter two. We evaluate this expression using Wick's theorem which only gives the following inelastic diagramatic contraction: $\braket{c_k^\dagger c_{k'-q'}}\braket{c_{k-q} c_{k'}^\dagger} $. The propagator for the free field theory is $\braket{c_k^\dagger c_{k'}} = \delta_{kk'} f(k)$ where $f(k)$ is the Fermi function evaluated at $\epsilon_k$. The above expression reduces to
\begin{equation}
    \sum_k e^{i(\epsilon_k - \epsilon_{k-q})t} V_{kq} V_{k-q,-q} f(k)(1-f(k-q))
\end{equation}
Taking the continuum limit in $k$ and Fourier transforming in time gives the dynamical correlation function
\begin{widetext}
\begin{equation}\label{eqn:twobodychi}
    \chi^{MF}(q,\omega) = \int \frac{dk}{2\pi} \delta(\omega + \epsilon_k - \epsilon_{k-q}) V_{kq} V_{k-q,-q} f(k)(1-f(k-q))
\end{equation}    
\end{widetext}
where
\begin{equation}
    V_{kq} V_{k-q,-q} = \left(\cos \left(k-\frac{q}{2}\right)-\alpha \cos \left(\frac{q}{2}\right)\right)^2
\end{equation}
%
%
Using a dispersion $\epsilon(k) = - t \cos(k)$, the delta function can be rewritten as 
\begin{equation}
   \delta(\omega + \epsilon_k - \epsilon_{k-q}) = \sum_{k_0} \frac{\delta(k-k_0)}{\sqrt{(2t\sin(q/2))^2 - \omega^2}}
\end{equation}
where the sum is over the wavevectors $k_0$ that satisfy conservation of energy $\omega + \epsilon_{k_0} - \epsilon_{k_0 - q} = 0$, i.e.\ at low frequencies $k_0$ is $\pm k_F$. 
Note the identity $2 \sin \left(\frac{q}{2}\right) \sin \left(k-\frac{q}{2}\right)=\cos (k-q)-\cos (k)$. 
At $k_0$, we also have
\begin{equation}
    \cos^2(k_0 - q/2) = \frac{1}{(2t\sin(q/2))^2}\left[ (2t\sin(q/2))^2 - \omega^2 \right]
\end{equation}
This allows the result to be simplified, e.g.  for the case $\alpha=0$, i.e. XY Hamiltonian and Raman operator, for which
\begin{widetext}
\begin{equation}
    \chi^{MF}(q, \omega) \propto \sum_{k_0} \frac{\sqrt{ (2v_s \sin(q/2))^2 - \omega^2 }}{(2v_s \sin(q/2))^2} f(k_0)(1- f(k_0 - q))
\end{equation}    
\end{widetext}
For $q = \pi$ and any $\alpha$, we recover \eqnref{eqn:cleanmf}.

\section{Linearized Mean Field Theory}\label{sec:linearmft}
Here again for completeness we walk an interested reader through the standard manipulations for computing the low energy fermion response function \eqnref{eqn:twobodychi}. At low energies, the response is proportional to the density-density correlation function given by
\begin{equation}
    \chi^{MF}(q,\omega) \approx W(q,\omega) \chi_{\rho\rho}''(q,\omega)
\end{equation}
where $W$ is an overall function $W$ of $q$ and $\omega$ and the density-density correlation function is 
\begin{equation}\label{eqn:chirhorho}
    \chi_{\rho\rho}''(q,\omega) = \int dt \ e^{i\omega t} \braket{\rho_{q}^\dagger (t) \rho_{q}(0)}_0
\end{equation}
with $\rho_q = \sum_k c_k^\dagger c_{k+q}$ and $\rho_q^\dagger = \rho_{-q}$. 

In the limit where $V_{kq} = 1$, \eqnref{eqn:twobodychi} is exactly the density-density correlation function. We thus first inspect the low energy response of \eqnref{eqn:chirhorho}. In this limit, the dispersion near the Fermi points is given by $\epsilon_k \approx \pm v k$. For $\omega$ small, energy conservation imposes $\epsilon_{k-q} \approx \epsilon_{k}$. Excitations at the same fermi point give $(\omega/v)\delta(\omega - v|q|)$, so we must look for excitations across both Fermi points. Suppose $k$ is near one of the Fermi points, say $k = k_F + \delta k$ with $|\delta k|$ small. For $q = 2k_F + \delta q$ with $|\delta q|$ small, we have $k-q = -k_F + \delta k - \delta q$ near the left Fermi point. For these excitations, we have $\omega = \epsilon_{k-q} - \epsilon_k = -v(\delta k - \delta q) - (v)(\delta k) = -v(2\delta k - \delta q)$. Rearranging, we have $\delta k = \frac12(\delta q - \omega/v)$ or $k = k_F + \frac12(\delta q - \omega/v)$. Similarly, $k-q = -k_F + \frac12(-\delta q - \omega/v)$. The Fermi functions place $k$ below the right Fermi point $(|k| < k_F)$ and $k-q$ above the left Fermi point ($|k-q| > k_F$). These constraints give $\delta q - \omega/v < 0$ and $-\delta q - \omega/v < 0$. Together, we have $\omega > v|\delta q| = v|q - 2k_F|$. A similar analysis of excitations from the left Fermi point to the right Fermi point gives $\omega > v|q + 2k_F|$. 

Integration of \eqnref{eqn:twobodychi} over $k$ yields a proportionality factor of the product of the vertices $V_{kq}V_{k-q,-q}$, with $k = \pm \frac12 (q - \omega/v)$. The product of the vertices thus becomes an overall function of $q$ and $\omega$ given by
\begin{equation}
    W(q,\omega) = V_{\pm \frac12 (q - \omega/v), q} V_{\mp \frac12(q+\omega/v), -q}
\end{equation}
The vertex functions evaluate to an overall function $W$ of $q$ and $\omega$. The low energy response of $\eqnref{eqn:twobodychi}$ is given by
\begin{equation}
    \chi^{MF}(q,\omega) \sim W(q,\omega) \chi_{\rho\rho}''(q,\omega)
\end{equation}
where the low energy response of $\chi_{\rho\rho}''$ is 
\begin{equation}
    \chi_{\rho\rho}''(q, \omega) \sim  \sum_\pm \Theta(\omega - v|q\pm 2k_F|) + (\omega/v) \delta(\omega - v|q|) 
\end{equation}    
with $\Theta$ being the Heaviside step function. 
The low energy response of $\chi^{MF}$ thus follows the low energy response of $\chi_{\rho\rho}''$.

\section{Bosonization}\label{sec:appbos}

We incorporate interactions beyond mean field to capture the low energy Raman response of zigzag domain walls in  bosonization. Recall the Luttinger liquid action which describes the low energy physics of the 1D $XXZ$ model \cite{giamarchi2004} 
\begin{equation}\label{eqn:llaction}
    S_{\text{Luttinger}} = \frac{1}{2\pi K} \int dx \ d\tau \left[ \frac{1}{v_s}(\partial_\tau \phi)^2 + v_s(\partial_x\phi)^2 \right]
\end{equation}
which is fully parameterized by the spinon velocity $v_s$ and the Luttinger parameter $K$. In our notation, at $K = 1$ the theory corresponds to the U(1) point (isotropic $XY$ model) and at $K = 1/2$ it corresponds to the SU(2) symmetric point (isotropic Heisenberg model). In the $K \to 1$ limit, mean field analysis becomes exact. 

While the Luttinger parameter $K$ generically depends on the applied magnetic field, in small magnetic fields we consider the case where $K$ is not appreciably renormalized. In applied magnetic field ($h^z > 0$) $K$ is renormalized away from $K = 1/2$: as $h^z$ approaches the saturation field $h_{sat} = 2J$, $K$ approaches the free point $K = 1$ \cite{cabra2004}. However for small magnetic fields near the critical field $h_c = \frac12 v_s (\pi/L_d)$ where $L_d$ is the length of a zigzag domain, we may qualitatively understand the physics at $K = 1/2$ without renormalization.

In order to capture the effect of small applied magnetic fields, we track how the wavevector of the Fermi points shifts due to the chemical potential of fermionic spinons created by a nonzero magnetic field. Recall that bosonization is the low energy field theory at the Fermi points, but the location (momenta) of the Fermi points is an external parameter that does not appear within the bosonized theory. Explicitly, 
\begin{equation}\label{eqn:chibosh}
\chi^{bos}(q, \omega, h^z) = \chi_{\rho\rho}(q-2h^z/v_s,\omega)
\end{equation}
We capture the effect of  domain walls by consider a Raman response at a shifted wavevector of $q = \pi \pm \pi/L_d$ with a zigzag domain length of $L_d$, or a distribution of domain sizes.

\section{Effect of Zigzag Domain Walls on the Hamiltonian}
Domain walls have a local effect in two ways. Modifications in the Hamiltonian would manifest as changes in the spin exchange coupling due to any changes in the bond length. The relative orientation of these bonds with photon polarizations would also affect their bond strength in the Raman operator. 

\begin{figure*}
    \centering
    \includegraphics{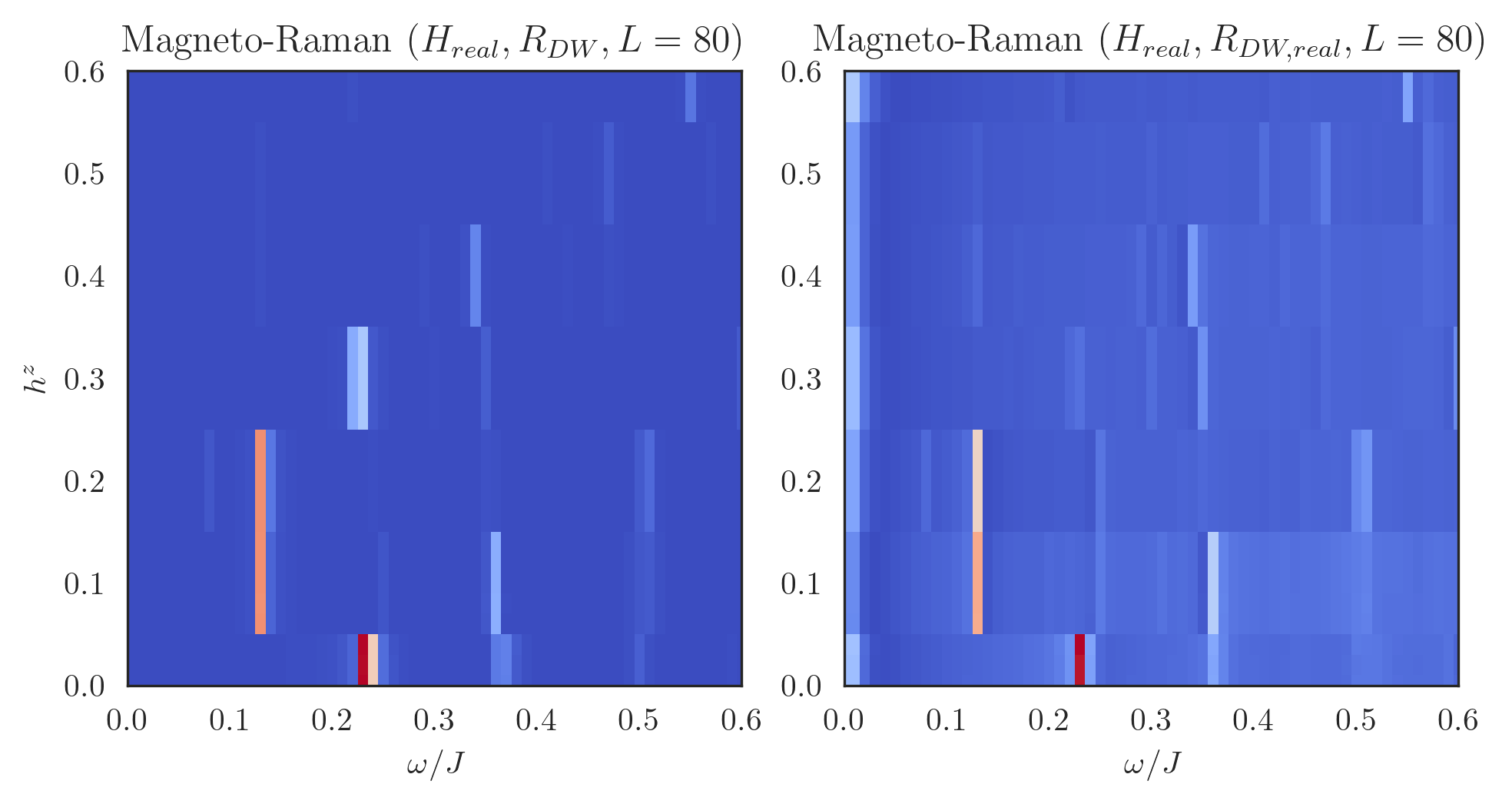}
    \caption{Numerically computed inelastic magneto-Raman spectra for finite sized systems ($L {=} 80$ sites) where zigzag domain walls modify the Hamiltonian (color scale is identical to Fig.~\ref{fig:tebd}). In both cases, the system hosts zigzag domain walls on lattice sites $j_1 {=} 26$ and $j_2 {=} 54$. The Hamiltonian is $H_\text{real} = \sum_{j=1}^{L} J_j \mathbf{S}_j \cdot \mathbf{S}_{j+1}$ with $J_j = 1$ for $j \neq j_1, j_2$ and $J_j = 0.3$ for $j = j_1, j_2$. The Raman operator $R_{DW}$ is the same as Fig.~\ref{fig:tebd}, while the Raman operator $R_{DW,\text{real}}$ is defined similarly to $H_\text{real}$. Explicitly $R_{DW,\text{real}} = \sum_{j=1}^L g_j \mathbf{S}_j \cdot \mathbf{S}_{j+1}$ with $g_j = \frac12 (1 + (-1)^j)$ for $j < j_1$ and $j > j_2$, $g_j = \frac12 (1 + (-1)^{j+1})$ for $j_1 < j < j_2$, and $g_j = 0.5$ for $j = j_1, j_2$. 
    (Left) Magneto-Raman spectrum when zigzag domain walls only modify the Hamiltonian. (Right) Magneto-Raman spectrum when zigzag domain walls modify both the Hamiltonian and the Raman operator. In both cases, we obtain the same qualitative features as Fig.~\ref{fig:tebd}.}
    \label{fig:realcolorplots}
\end{figure*}

Even when domain walls appreciably modify the Hamiltonian or the Raman operator, we find no qualitative change in our conclusions. We may capture such changes directly in our numerics. When only spin exchange couplings in the Hamiltonian are modified, the magneto-Raman spectrum (Fig.~\ref{fig:realcolorplots}, left) qualitatively reproduces the spectrum from the main text (Fig.~\ref{fig:tebd}). When both the Hamiltonian and the Raman operator are modified by zigzag domain walls, we again find the same qualitative spectral features (Fig.~\ref{fig:realcolorplots}, right).

Local modifications to the Raman operator add disorder which can appear as background noise, but don't do much else. In particular, the qualitative features of the Raman response may still be captured by the most dominant Fourier mode $q^*$ of the Raman operator by studying $\chi(q^*, \omega)$. 

Changes to the Hamiltonian, however, are more subtle. We may consider two limiting cases where domain walls do or don't modify the Hamiltonian. In one limit, the Hamiltonian is not modified at all. In this case as discussed in the main text, we may approximate the Raman response by consider a Raman operator with a smoothed out bond profile which has nodes precisely at the defects.
In the opposite limit, we may consider the case where domain walls do modify the Hamiltonian locally, and consider the 1D case (as in the current toy model) where such local perturbations are RG relevant and flow to infinite strength. The fixed point of the renormalization group flow takes the system to an open chain. This fragmented chains limit is discussed in the main text.

\section{Raman in Linear Spin Wave Theory}
In the main text, we demonstrated an anomalous, singular response to magnetic fields that may be observed by Raman scattering in phases without magnetic order, but we did not consider whether such a response could be measured in magnetically ordered phases. For completeness, we here study in detail whether any such responses arise in magnetic order, and find that they indeed do not.

For magnetically ordered phases  spin excitations from the ground state are described by magnons. In this section, we compute the Raman response of $R_q$ \eqnref{eqn:rq} using magnons for the 1D Heisenberg Hamiltonian in applied field. We compute this response within linear spin-wave theory (LSWT) and only to lowest non-vanishing order in bosonic operators.

It is well known \cite{mermin1966}, however, that quantum fluctuations in 1D diverge and obstruct the precipitation of magnetic order. Nonetheless, we employ LSWT in order to gain intuition for how quasi-1D spin chains may behave in the presence of magnetic order mediated by interactions in 3D. We find that bosonic magnon excitations, unlike fermionic spinon excitations, lack the anomalous, singular magnetic field response we present in the main text. Whereas an applied magnetic field shifts the Fermi momentum of a spinon Fermi surface, and may tune gapped excitations to become gapless, bosonic magnon excitations do not experience such an effect. Indeed, for 1D magnons we find no similar anomalous magnetic field response that is probed by Raman scattering. 

To begin, we first consider magnons in a ferromagnetic phase in applied magnetic field. We then consider the antiferromagnetic phase at zero applied field. In this case, we find the Raman response to be non-vanishing at the 4-magnon level. In applied field, however, the classical ground state of the antiferromagnet changes. At finite field, the Raman response then becomes non-vanishing at the 2-magnon level and mimics the dynamical structure factor probed by neutron scattering. In all cases, however, the magnetic field dependence of the Raman response lacks the anomaly we find for free spinons in the main text. 

\subsection{Ferromagnetic Phase}
For the spin-$S$ ferromagnetic Heisenberg Hamiltonian ($J < 0$) with periodic boundary conditions, we may transform spin operators to bosonic magnon operators via the Holstein-Primakoff transformation \eqnref{eqn:hp} 
\begin{equation}\label{eqn:hp}
    S_i^z = S - a_i^\dagger a_i, \ S_i^- = a_i^\dagger \sqrt{2S - a_i^\dagger a_i}, \ S_i^+ = (S_i^-)^\dagger 
\end{equation}
with $[a_i, a_j^\dagger] = \delta_{ij}$. Within LSWT, we expand in powers of $(a_i^\dagger a_i/2S)$ and keep bosonic operators up to quadratic order. For the present case, LSWT yields a free bosonic Hamiltonian \eqnref{eqn:fmhamiltonian} after a Fourier transform
\begin{equation}\label{eqn:fmhamiltonian}
    H_{FM} = \sum_k \epsilon_k a_k^\dagger a_k, \quad \epsilon_k = 2|J|S(1-\cos(k)) + h^z
\end{equation}
The Raman operator $R_q$ within LSWT reads
\begin{equation}
    R_q = \sum_k \Gamma_{kq} a_k^\dagger a_{k+q}
\end{equation}
with the vertex given by \eqnref{eqn:fmvertex}. 
\begin{equation}\label{eqn:fmvertex}
    \Gamma_{kq} = S(e^{-i(k+q)} + e^{ik} - 1 - e^{-iq})
\end{equation}
The Raman operator $R_q$ is thus a magnon density excitation at momentum $q$. To lowest order in bosonic operators, the first non-vanishing contribution to the Raman response $\chi^{FM}$ is a 4-magnon response (at finite $T$). 

We perform calculations analogous to the main text and find
\begin{widetext}
\begin{equation}
    \chi^{FM}(q,\omega) = \sum_{k_0} \frac{\Gamma_{k_0, q} \Gamma_{k_0+q, -q} n(k_0)(1+n(k_0 + q))}{\sqrt{(4JS\sin(q/2))^2 - \omega^2}} 
\end{equation}    
\end{widetext}
where $k_0 \in [-\pi,\pi]$ satisfies $\omega + \epsilon_{k_0} - \epsilon_{k_0 + q} = 0$, and Bose functions are given by $n(k) = (e^{\beta\epsilon_k} - 1)^{-1}$. 

At $T = 0$, however, the ground state is the magnon vacuum. The Raman response at this order vanishes since $R_q(t = 0)$ annihilates the ground state. As such we conclude the magnon-Raman response of $R_q$ vanishes within LSWT for the ferromagnetic phase. 

\subsection{Antiferromagnetic Phase ($h = 0$)}\label{sec:afmlswt}
At zero magnetic field, the classical ground state of the antiferromagnetic phase is the N{\'e}el state. To perform LSWT, we first apply rotate spins by $\pi$ about the $x$-axis on every other lattice site. Under this transformation, spins on one sublattice transform like $S_j^z \mapsto -S_j^z$ and $S_j^\pm \mapsto S_j^\mp$. Under this rotation, Holstein-Primakoff bosons are the appropriate fluctuations about the classical antiferromagnetic ground state. The Heisenberg Hamiltonian after this sublattice rotation is given by \eqnref{eqn:afmheisrot}. 
\begin{equation}\label{eqn:afmheisrot}
    H = J\sum_j \frac12\left( S_j^- S_{j+1}^- + S_j^+ S_{j+1}^+ \right) - S_j^z S_{j+1}^z
\end{equation}
Keeping only quadratic terms in Holstein-Primakoff bosons and performing a Fourier transform, the Hamiltonian becomes (up to an overall constant)
\begin{equation}
    H = 2JS\sum_k a_k^\dagger a_k + \frac12 \gamma_k (a_{k}a_{-k} + a_{k}^\dagger a_{-k}^\dagger )
\end{equation}
where $\gamma_k = \cos(k)$ is the sum of lattice harmonics in 1D. While the Hamiltonian is not diagonal in these bosonic operators, we may employ a Bogoliubov transform to diagonalize it
\begin{equation}
    \begin{pmatrix}
     b_k \\ b_{-k}^\dagger
     \end{pmatrix} = 
     \begin{pmatrix}
     \cosh\phi_k & \sinh\phi_k \\
     \sinh\phi_k & \cosh\phi_k
     \end{pmatrix}
     \begin{pmatrix}
         a_k \\ a_{-k}^\dagger
     \end{pmatrix}
\end{equation}
The Hamiltonian is diagonalized for $\gamma_k = \tanh2\phi_k$ and reads
\begin{equation}
    H = \sum_k \epsilon_k (b_k^\dagger b_k + 1/2), \quad \epsilon_k = 2JS|\sin(k)|
\end{equation}
with $[b_k, b_{k'}^\dagger] = \delta_{kk'}$. 

The Raman operator $R_q$ within LSWT may be similarly constructed. We find
\begin{equation}
    R_q = \sum_k \Psi_k^\dagger M_{kq} \Psi_{k+q}, \quad \Psi_k = 
    \begin{pmatrix}
        b_k \\ b_{-k}^\dagger
    \end{pmatrix}
\end{equation}
with the vertex given by \eqnref{eqn:afmmagnonvertex}. In the expression, $\sigma^0, \sigma^1$ denote the $2\times 2$ identity and $x$ Pauli matrices respectively.
\begin{widetext}
\begin{equation}\label{eqn:afmmagnonvertex}
    M_{kq} = \sigma^0|\csc(k)|\left[ \frac12(1+e^{iq}) - e^{-i(k+q)}\cos(k) \right] + \sigma^1|\csc(k)|\left[ e^{-i(k+q)} - \frac12(1+e^{iq})\cos(k) \right]
\end{equation}
\end{widetext}
Time evolution of $R_q$ in this form is straightforward and given by 
\begin{equation}
    R_q(t) = \sum_k \Psi_k^\dagger U_k^\dagger(t) M_{kq} U_{k+q}(t) \Psi_{k+q}
\end{equation}
where $U_k(t) = \exp(-i\sigma^3\epsilon_k t)$ and $\sigma^3$ is the usual $z$ Pauli matrix. 

The Raman response may then be evaluated diagrammatically. At $T = 0$, the ground state of the system is the magnon vacuum. Hence only correlations of the form $\braket{a_{k_1}(t) a_{k_2}(t) a_{k_3}^\dagger(0) a_{k_4}^\dagger(0)}$ contribute to a finite response. In total, we find a finite response 
\eqnref{eqn:afmzerofieldresponse} which follows the magnon dispersion. 
\begin{widetext}
\begin{equation}\label{eqn:afmzerofieldresponse}
    \chi^{AFM}(q, \omega) = \frac{1}{\sqrt{(4JS\cos(q/2))^2 - \omega^2}} \sum_{k_0 \in  [0,\pi], k_0 \in [-\pi, 0]} (M_{k_0,q})_{21} \left[ (M_{-k_0,-q})_{12} + (M_{k_0+q,-q})_{12} \right]
\end{equation}
\end{widetext}
In the response \eqnref{eqn:afmzerofieldresponse}, the sum over $k_0 \in [0,\pi],[-\pi,0]$ is taken over the right/left halves of the first Brillouin zone, with $k_0$ satisfying $\omega - (\epsilon_{k_0} + \epsilon_{k_0+q}) = 0$, and $(M_{kq})_{ij}$ denote the $ij$-th matrix element of the vertex expressed in \eqnref{eqn:afmmagnonvertex}.

\subsection{Antiferromagnetic Phase ($h > 0$)}
While at zero magnetic field, the classical ground state of the antiferromagnetic phase is the N{\'e}el state, in applied magnetic field this is not the case. Classical spins cant by an angle $\theta_c$ to align with the applied field. Following \cite{zhitomirsky1998,zhitomirsky1999,syljuasen2008}, we allow spins to cant by an angle $\theta$ about the $\hat{y}$ axis away from the N{\'e}el state. The classical ground state energy is minimized for $\sin\theta_c = h^z/4JS$. 

Within LSWT, the Hamiltonian no longer conserves magnon number. Indeed, we find
\begin{equation}
    H = \sum_k A_k a_k^\dagger a_k - \frac{B_k}{2}(a_k a_{-k} + a_{k}^\dagger a_{-k}^\dagger)
\end{equation}
with $A_k = 2JS(1 + \sin^2(\theta_c) \gamma_k)$, $B_k = 2JS\cos^2(\theta_c) \gamma_k$, and $\gamma_k = \cos(k)$ is the sum of lattice harmonics for the 1D spin chain. Note, for $h = 0 \implies \theta_c = 0$, the coefficients $A_k, B_k$ are precisely those found in the previous section. We may again employ a Bogoliubov transformation to diagonalize $H$. In this case, the Hamiltonian is diagonalized for $\tanh(2\phi_k) = -B_k/A_k$. Then
\begin{equation}
    H = \sum_k \epsilon_k (b_k^\dagger b_k + 1/2)
\end{equation}
where
\begin{equation}
    \epsilon_k = 2JS\sqrt{(1+\gamma_k)(1-\cos(2\theta_c)\gamma_k)}
\end{equation}
For $\theta = \theta_c$, the linear magnon terms in $H$ vanish identically. For the Raman operator $R_q$, however, this is not the case. Indeed, to lowest order in Bogoliubov magnons $b_k$, we find
\begin{align}
    R_q = \sqrt{\frac{S^3}{2}} \sin(2\theta_c)(1 {+} e^{-iq}) e^{\phi_{-q+\pi}}\left( 
    b_{q+\pi} {+} b_{-(q+\pi)}^\dagger \right)
\end{align}    
Physically, $R_q$ creates a single magnon excitation at $\pm(q+\pi)$. 

The Raman response $\chi^{AFM}$ of $R_q$ to lowest order is a 2-magnon response. At this order, we find
\begin{equation}
    \chi^{AFM}(q,\omega) = 2JS^4\sin(2\theta_c) \sin^2(q) \frac{\delta(\omega - \epsilon_{q+\pi})}{\omega}
\end{equation}
At $h^z = 0$, $\theta_c = 0$ and so this contribution of the response vanish. 

We may follow a similar procedure as in the ferromagnetic phase to compute the 4-magnon response. We find the response qualitatively follows the $h^z \to 0^+$ limit of the 2-magnon response, and so we omit the calculation here. 

\section{Raman response of geometric defects (small amplitude bond variation)}

In this section we will consider small amplitude bond variations in a zigzag chain and determine the subsequent Raman response. We will restrict ourselves to the nearest neighbor Heisenberg Hamiltonian $H = J\sum_j \mathbf{S}_j \cdot \mathbf{S}_{j+1}$. Suppose the bond angle generically depends on the lattice site
\begin{equation}
    \theta_j = \theta_0 + \epsilon f_j
\end{equation}
where $\theta_0$ is the equilibrium bond angle and $\epsilon \ll 1$. We let the variation $f$ in the bond angle be arbitrary.
\begin{equation}
    R = \sum_j (\ehat_i \cdot \mathbf{r}_{j})(\ehat_s \cdot \mathbf{r}_{j}) (\mathbf{S}_j \cdot \mathbf{S}_{j+1})
\end{equation}
where $\mathbf{r}_j$ is the bond vector pointing from site $j$ to site $j+1$. For a zigzag chain formed from isosceles triangles with reflection symmetry identifying zig and zag bonds, we have
\begin{equation}
    \mathbf{r}_j = \begin{pmatrix}
    \cos(\theta_j) \\ (-1)^j \sin(\theta_j)
    \end{pmatrix}, \quad \ehat_{i,s} = \begin{pmatrix}
    \cos\theta_{i,s} \\ \sin\theta_{i,s}
    \end{pmatrix}
\end{equation}
When $\theta_i + \theta_s = \frac{\pi}{2}$, to order $\textit{O}(\epsilon^2)$ we find (up to spectral equivalence)
\begin{align}
    (\ehat_i \cdot \mathbf{r}_{j})(\ehat_s \cdot \mathbf{r}_{j}) = (-1)^j &\big[ \frac12 \sin(2\theta_0) + \epsilon f(j) \cos(2\theta_0) \nonumber
    \\ &
    - (\epsilon f(j))^2 \sin(2\theta_0) \big]
\end{align}    
We next define the following operators
\begin{align}
    R_q &= \sum_j e^{iqj} (\textbf{S}_j \cdot \textbf{S}_{j+1}) \\
    R[f] &= \sum_j (-1)^j f_j (\textbf{S}_j \cdot \textbf{S}_{j+1}) = \sum_q \tilde{f}_{q+\pi} R_q
\end{align}
where $\tilde{f}_q$ is the Fourier transform of $f_j$. The Raman operator then reads
\begin{align}
    R = & \frac12 \sin(2\theta_0) R_\pi + \epsilon \cos(2\theta_0) \sum_q \tilde{f}_{q+\pi} R_q 
    \nonumber \\ &
    + \epsilon^2 (-\sin(2\theta_0)) \sum_q \tilde{(f^2)}_{q+\pi} R_q
\end{align}
We then compute the Raman correlator $\braket{R_q(t) R_{q'}(0)}$ to find the Raman response $\chi(q, \omega)$ as before. The Raman response with small angle variations is then readily found to be $I(\omega) = I_0(\omega) + \epsilon I_1(\omega) + \epsilon^2 I_2(\omega) + O(\epsilon^3)$
where each intensity contribution is given by
\begin{widetext}
\begin{align}
    I_0(\omega) &= \frac14 \sin^2(2\theta_0) \chi(\pi, \omega) \\
    I_1(\omega) &= \sin(2\theta_0)\cos(2\theta_0) \sum_q \tilde{f}_{q+\pi} \int dt \ e^{i\omega t} \braket{R_\pi(t) R_q(0)} \nonumber \\ 
    &= \sin(2\theta_0)\cos(2\theta_0) \tilde{f}_0 \chi(\pi, \omega) \\
    I_2(\omega) &= -\sin^2(2\theta_0) \sum_q \tilde{(f^2)}_{q+\pi} \int dt \ e^{i\omega t} \braket{R_\pi(t) R_q(0)} + \cos^2(2\theta_0)\sum_{qq'} \tilde{f}_{q+\pi}\tilde{f}_{q'+\pi} \int dt \ e^{i\omega t}\braket{R_q(t) R_{q'}(0)} \nonumber \\
    &= -\sin^2(2\theta_0) \tilde{(f^2)}_{0} \chi(\pi, \omega)  + \cos^2(2\theta_0) \sum_q |\tilde{f}_{q}|^2 \chi(q+\pi, \omega)
    \\
    I(\omega) 
    &= \sin(2\theta_0) \left( \frac12\sin(2\theta_0) + \epsilon \cos(2\theta_0)\tilde{f}_0 - \epsilon^2 \sin(2\theta_0) \tilde{(f^2)}_{0} \right)\chi(\pi, \omega) + \epsilon^2 \cos(2\theta_0)\sum_q |\tilde{f}_q|^2 \chi(q+\pi, \omega)
    \label{eqn:geoepsilon}
\end{align}    
\end{widetext}

\section{Mathematical treatment of Poisson distributions}
The power spectrum of the Raman operator bond profile can be computed exactly when zigzag domain sizes are Poisson distributed -- an unphysical limit but worth discussing for mathematical completeness. Consider a random telegraph signal generated by a Poisson point process. 

Let $g_j$ take on values of $\pm 1$. Suppose the probability of observing $m$ zigzag domains with positive parity over a distance $x$ is given by
\begin{equation}
    p(m, x) = \frac{(\mu x)^m}{m!} e^{-\mu x}
\end{equation}
with $\mu$ the mean number of positive parity domains per unit length. We may then compute the autocorrelation function $\phi_x(g) = g_j g_{j+x}$. The autocorrelation function $\phi_x(g)$ is $+1$ if the number of domains with particular parity found in the interval of lattice sites $(j, j+x)$ is even. Similarly $\phi_x(g) = -1$ if the number is odd. Hence the autocorrelation is given by  
\begin{equation}
    \phi_x(g) = \sum_{k=0}^\infty p(2k, x) - p(2k+1, x)= e^{-2\mu x} 
\end{equation}

The power spectrum of $g_j$ is then the Fourier transform of $\phi_x(g)$ by the Wiener-Khinchin theorem.
This is a Lorentzian centered at $q = 0$ whose full-width at half-maximum scales linearly with $\mu^{-1}$. 

\begin{equation}\label{eqn:ftpoisson}
    |g_q|^2 \propto \frac{1/\mu}{(1+(q/2\mu)^2}
\end{equation}


\nocite{*}

\bibliography{apssamp}

\end{document}